\documentclass[3p]{elsarticle} 

\usepackage{lineno,hyperref}
\modulolinenumbers[5]
\usepackage{changepage}
\journal{Corrosion Science}

\usepackage{amsmath} 
\usepackage{amssymb}
\usepackage{bm} 
\usepackage{mathtools} 
\biboptions{sort&compress}
\usepackage{makecell}  
\usepackage{float} 
\usepackage{tablefootnote}
\usepackage{subcaption} 
\usepackage{placeins}

\usepackage{xcolor,cancel}
\usepackage{color,soul}
\usepackage{calc}
\newsavebox\CBox
\newcommand\hcancel[2][0.5pt]{%
  \ifmmode\sbox\CBox{$#2$}\else\sbox\CBox{#2}\fi%
  \makebox[0pt][l]{\usebox\CBox}%
  \rule[0.5\ht\CBox-#1/2]{\wd\CBox}{#1}}

\begin{document}

\begin{frontmatter}
\title{An electro-chemo-mechanical framework for predicting hydrogen uptake in metals due to aqueous electrolytes}

\author{Tim Hageman}
\author{Emilio Martínez-Pañeda \corref{mycorrespondingauthor}}
\cortext[mycorrespondingauthor]{Corresponding author}
\ead{e.martinez-paneda@imperial.ac.uk}

\address{Department of Civil and Environmental Engineering, Imperial College London, London SW7 2AZ, UK}

\begin{abstract}
We present a theoretical and numerical scheme that enables quantifying hydrogen ingress in metals for arbitrary environments and defect geometries. This is achieved by explicitly resolving the electrochemical behaviour of the electrolyte, the hydrogen and corrosion reactions, the kinetics of surface adsorption, and hydrogen uptake, diffusion and trapping in mechanically-deforming solids. This new framework is used to produce maps that relate the absorbed hydrogen with the applied potential, specimen geometry and fluid velocity. We also present simplified versions of our generalised model, and benchmark predictions of these and other existing models against the generalised electro-chemo-mechanical results, establishing regimes of validity.  

\end{abstract}


\end{frontmatter}


\section{Introduction}

The ingress of hydrogen into a metal brings a reduction in material toughness, ductility and fatigue crack growth resistance \cite{Gangloff2003,Gangloff2012,Djukic2019}. This phenomenon, often referred to as \emph{hydrogen embrittlement}, is pervasive across the energy, defence, transport and construction sectors, and is gaining increasing attention due to the higher susceptibility of modern, high-strength alloys \cite{RILEM2021}. As a result, there is a significant body of literature devoted to the development of chemo-mechanical models for predicting hydrogen assisted failures (see, e.g., \cite{Yu2016a,Nagao2018,CMAME2018,Anand2019,Shishvan2020,IJP2021} and Refs. therein). These hydrogen embrittlement models commonly solve a coupled deformation-diffusion problem to define a fracture criterion as a function of mechanical fields (stress, strain) and hydrogen concentration. Consistent with experimental observations, model predictions are very sensitive to the hydrogen content, a key input in the transport sub-problem. However, the quantification of hydrogen ingress remains a challenge, and this is particularly the case when hydrogen originates from water vapour or aqueous electrolytes \cite{Marcus2012,Turnbull2015}. In other words, our inability to quantify hydrogen uptake is holding back the predictive potential of current hydrogen embrittlement models.\\ 

By far, the most widely used strategy for modelling hydrogen ingress is the definition of a constant hydrogen concentration at the surfaces of the sample exposed to the hydrogen-containing environment (see, e.g., \cite{Yu2016a,Nagao2018,CMAME2018,IJP2021,Moriconi2014,Duda2018,CS2020b,Wu2020b,Colombo2020}). Recently, a few authors have proposed instead to prescribe a constant chemical potential \cite{DiLeo2013,IJHE2016,Diaz2016b,Elmukashfi2020,AM2020}. This is more accurate as it enables capturing the increase in hydrogen solubility associated with volumetric strains (lattice dilatation effects near the surface). However, neither the hydrogen content nor the chemical potential are typically known in most hydrogen-containing environments, such as aqueous electrolytes. Atrens and co-workers \cite{Liu2014,Venezuela2018a} postulated the concept of an equivalent fugacity to relate the hydrogen content at the surface to the overpotential through experimental calibration. Turnbull and co-workers \cite{Turnbull1996,CS2020} and Kehler and Scully \cite{Kehler2008} have gone one step further and, building upon a number of assumptions, respectively defined the flux and the hydrogen concentration to be a function of the absorption and desorption reaction rate constants. In the regimes where their assumptions are relevant, this enables quantifying hydrogen ingress as a function of the overpotential and pH of the environment. However, the \emph{local} pH and overpotential are typically unknown and can differ significantly from the \emph{bulk} pH and overpotential, the commonly known quantities. For example, the narrow confines of occluded areas such as cracks or pits limit the exchange of dissolved metal ions with the bulk electrolyte, resulting in a very different local chemistry – e.g., the pH can change from 9 (global) to 2 (local) \cite{Mccafferty2004,Carneiro-Neto2016,Duddu2016}. In turn, these pH differences can result in different reaction rates \citep{Recio2011,Fujimoto2017}, and can thus result in significant differences in absorbed hydrogen between the area near the crack compared to the exterior boundaries \citep{Cooper2007}. An accurate estimation of hydrogen ingress requires resolving not only the absorption kinetics but also the bulk and surface electrochemistries, coupled with bulk hydrogen diffusion and mechanical straining.\\ 

In this work, we present a theoretical and computational modelling framework that fully resolves the physics of hydrogen uptake. The model combines: (i) the electrochemical behaviour of the electrolyte (ion transport, electrolyte potential distribution), (ii) the Volmer, Heyrovsky, and Tafel reactions intrinsic to the Hydrogen Evolution Reaction (HER), (iii) adsorption and absorption surface kinetics, and (iv) hydrogen ingress, diffusion and trapping in a mechanically-deforming solid. For the first time, the electrochemistry of hydrogen uptake is explicitly modelled, enabling us to establish a connection between the bulk environment and the influx of hydrogen for arbitrary sample and defect geometries. Moreover, unlike previous attempts to connect the environment to the hydrogen ingress process, we do not establish any \textit{a priori} assumptions and thus do not limit our predictions to specific conditions. The competition between different reaction rates is investigated as a function of the applied electric potential and pH, establishing regimes of dominance and particularising the generalised model presented. Our calculations span a wide range of applied potentials, from anodic to cathodic regimes, mapping the impact of the environment on hydrogen absorption. Furthermore, we study the influence of the crack geometry and fluid flow velocity, establishing the scenarios where these effects are important. Finally, we provide simplified models which, together with the maps provided, can be used to approximate hydrogen ingress without explicitly simulating the electrolyte. The performance of these simplified models is compared to the results obtained from the complete electro-chemo-mechanical framework as well as to those calculated employing commonly used boundary conditions.\\

The remainder of this paper is structured as follows. The theory and governing equations are presented in Section \ref{sec:gov_eq}. The numerical framework is then briefly described in Section \ref{sec:disc}. Subsequently, model predictions are validated against computational and experimental results in Section \ref{sec:verif}. In Section \ref{sec:results}, hydrogen uptake is quantified as a function of the environmental conditions and the defect geometry. Finally, Section \ref{sec:bcs} shows how the results estimated with the complete and simplified models presented compare with commonly used modelling strategies. Concluding remarks end the manuscript in Section \ref{sec:conclusion}. 

\section{Theory}
\label{sec:gov_eq}

\begin{figure}
    \centering
    \includegraphics{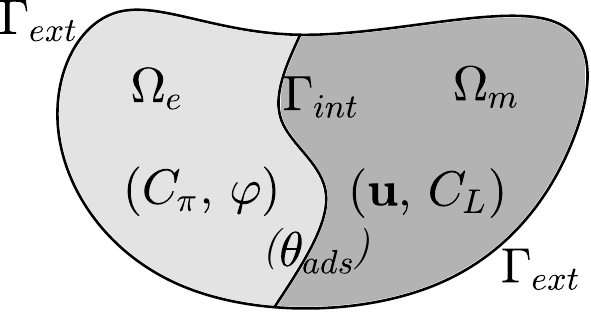}
    \caption{Schematic overview of the electrolyte and metal domains, and (in brackets) the degrees of freedom used to describe the electrochemical phenomena relevant to each of them.}
    \label{fig:domains}
\end{figure}

We consider a domain composed of two parts, an electrolyte in $\Omega_e$ and a metal in $\Omega_m$, as shown in Fig. \ref{fig:domains}. These two sub-domains interact on the interface $\Gamma_{int}$. In terms of primary fields, the metal domain is described through its displacement vector $\mathbf{u}$ and the hydrogen concentration at interstitial lattice sites $C_L$; the electrolyte behaviour is characterised by the concentration of the $\pi$ ionic species $C_\pi$ and the electric potential $\varphi$; and the internal interface is described through the coverage of adsorbed hydrogen $\theta_{ads}$. The focus is on surface reactions and hydrogen diffusion, and consequently no crack growth or material dissolution is considered, with pits and cracks being represented geometrically through the shape of the simulated domains.\\ 

A note on notation and units. For consistency across domains, the SI units $\mathrm{mol}/\mathrm{m}^3$ are used for all concentration quantities. As a consequence, reaction constants units follow accordingly; e.g., the water auto-ionization constant is given by $10^{-8}\;(\mathrm{mol}/\mathrm{m^3})^2$ (as opposed to the more common terminology of $10^{-14}\;(\mathrm{mol}/\mathrm{L})^2$). Also, we define the electric potential relative to the standard hydrogen electrode, considering the equilibrium potential of hydrogen related reactions to be at $0\;\mathrm{V}_{SHE}$. We use lightface italic letters for scalars, e.g. $C_L$, upright bold letters for vectors, e.g. $\mathbf{u}$, and bold italic letters, such as $\bm{\sigma}$, for second and higher order tensors. First-, second-, and fourth-order tensors are in most cases respectively represented by small Latin, small Greek, and capital Latin letters. The gradient and the divergence are respectively denoted by $\bm{\nabla}\mathbf{u}= u_{i,j}$ and $\bm{\nabla}\cdot\bm{\sigma}=\sigma_{ij,j}$. And the trace of a second order tensor is written as $\text{tr} \,\bm{\varepsilon}=\varepsilon_{ii}$. 

\subsection{Metal sub-domain}
\label{Sec:MetalDomain}

The transport of ions inside the electrolyte and of hydrogen inside the metal are much slower than the deformation experienced by the material. Accordingly, the metal can be assumed to be in a state of quasi-static equilibrium, while the diffusion of the ions and hydrogen atoms is time-dependent. This quasi-static mechanical equilibrium is characterised by the momentum balance: 
\begin{equation}
    \bm{\nabla}\cdot\bm{\sigma} = \mathbf{0}
    \label{eq:mombalance}
\end{equation}
in which the Cauchy stress tensor $\bm{\sigma}$ is estimated based on the assumption of linear-elastic material behaviour.\\ 

The hydrogen inside the metal is located at interstitial lattice sites, $C_L$, and within $i$ sets of hydrogen traps, $C_{T}^i$. The mass conservation for the total hydrogen content is given by (see, e.g. \cite{Dadfarnia2011,Fernandez-Sousa2022}):
\begin{equation}
    \dot{C}_L+\sum_i \dot{C}_T^i + \bm{\nabla}\cdot\left(-D_L \bm{\nabla}C_L \right) + \bm{\nabla}\cdot\left(\frac{D_L C_L \overline{V}_H}{RT}\bm{\nabla}\sigma_H\right) = 0
    \label{eq:massbalance1}
\end{equation}
where $T$ is the temperature, $R$ is the universal gas constant, and $\overline{V}_H$ is the partial molar volume of hydrogen in the material. We use $\dot{C_L}$ and $\dot{C}_T^i$ to indicate the time derivatives of the hydrogen concentrations within the lattice and traps (of type $i$). The equation assumes that hydrogen diffuses through the interstitial lattice sites with a diffusion coefficient $D_L$, whereas traps are isolated and do not form an extended network through which hydrogen atoms can diffuse. Additionally, there is a contribution from the hydrostatic stress $\sigma_H=\text{tr}(\bm{\sigma})/3$, as the hydrogen solubility increases in areas of high volumetric strains due to lattice dilatation.\\ 

Let us now define the relation between lattice and trapping sites, for which several models exist. The transfer of hydrogen atoms between lattice and trap sites can be simulated using a kinetic formulation \cite{McNabb1963,Turnbull1996,Turnbull1997}, which can capture trapping behaviour that is perfectly reversible, (quasi)-irreversible, and asymmetric (different absorption and desorption energies). Alternatively, a common assumption is that of fast trapping kinetics, which results in the consideration of all traps being reversible and results in an equilibrium relationship between the trapped and lattice hydrogen content \citep{Oriani1970,Diaz2019}. To formulate this equilibrium relationship, let us first introduce the densities of lattice ($N_L$) and trapping ($N_T^i$) sites, which allows then to define the occupancy of lattice and trapping sites as $\theta_L=C_L/N_L$ and $\theta_T^i=C_T^i/N_T^i$, respectively. Then, assuming a low lattice occupancy ($\theta_L<<1$), one can define the trap occupancy as a function of the lattice occupancy and the trap binding energy $E_b$, as follows:
\begin{equation}
    \theta_T^i = \frac{C_L/N_L \exp{\left(\frac{E_b^i}{RT}\right)}}{1+C_L/N_L \exp{\left(\frac{E_b^i}{RT}\right)}}
\end{equation}
This allows the mass balance (Eq. \eqref{eq:massbalance1}) to solely be given in terms of the lattice concentration as:
\begin{equation}
    \left(1+ \sum_i \frac{N_T^i/N_L \; \; \exp{\left(E_b^i/(RT)\right)}}{\left(1+C_L/N_L \exp{\left(E_b^i/(RT)\right)}\right)^2} \right)\dot{C}_L + \bm{\nabla}\cdot\left(-D_L \bm{\nabla}C_L \right) + \bm{\nabla}\cdot\left(\frac{D_L C_L \overline{V}_H}{RT}\bm{\nabla}\sigma_H\right) = 0
    \label{eq:massbalance2}
\end{equation}

Eqs. \eqref{eq:mombalance} and \eqref{eq:massbalance2} describe the mechanical behaviour of the metal and the transport of hydrogen within it. These equations are subject to the following boundary conditions on $\Gamma_{ext}$ and $\Gamma_{int}$:
\begin{align}
    \mathbf{u} = \overline{\mathbf{u}} \qquad &\text{or} \qquad \bm{\sigma}\cdot\mathbf{n} = \overline{\mathbf{t}} \\
    C_L = \overline{C}_L \qquad &\text{or} \qquad -D_L \bm{\nabla}C_L + \frac{D_L C_L \overline{V}_H}{RT}\bm{\nabla}\sigma_H = \overline{J}
    \label{eq:BCflux}
\end{align}
with $\overline{\mathbf{u}}$, $\overline{\mathbf{t}}$, $\overline{C}_L$, and $\overline{J}$ being the externally enforced displacements, tractions, lattice concentration, and hydrogen inflow flux. For the metal-electrolyte interface we assume a traction free boundary condition, $\overline{\mathbf{t}}=0$. Regarding the field $C_L$, a hydrogen influx $\overline{J}$ is generally defined at the interface, as detailed in Section \ref{Sec:ElectrolyteDomain} (Eq. \ref{eq:JfluxAbs}) but, for comparison purposes, results are also obtained with the simplistic and widely used boundary condition of prescribing a constant lattice hydrogen content $C_L$ (Section \ref{sec:bcs}).

\subsection{Electrolyte sub-domain}
\label{Sec:ElectrolyteDomain}

We model a seawater-like electrolyte, consisting of the ionic species $\mathrm{H}^+,\;\mathrm{OH}^-,\;\mathrm{Na}^+,\;\mathrm{Cl}^-$, and the dissolved metal concentration; here, iron $\mathrm{Fe}^{2+}$ and its reaction product $\mathrm{FeOH}^+$. Inside this electrolyte, each ionic species $\pi$ is described through their respective concentration $C_\pi$. In addition, an electric field $\varphi$ is present. The evolution of these concentrations is given through the Nernst-Planck mass balance:
\begin{equation}
    \dot{C}_{\pi}+(\mathbf{v}\cdot\bm{\nabla})c_\pi+\bm{\nabla}\cdot\left(-D_\pi \bm{\nabla}C_\pi\right) + \frac{z_\pi F}{RT} \bm{\nabla} \cdot \left(-D_\pi C_\pi \bm{\nabla} \varphi\right) +R_\pi = 0 \label{eq:nernstplanck}
\end{equation}
where $z_\pi$ is the ionic charge and $F$ is Faraday's constant. The velocity field of the electrolyte $\mathbf{v}$ is presumed to be known; i.e., electrolyte fluid flow simulations are not conducted. In addition to the mass balance, we assume electro-neutrality, requiring the electrolyte to be neutrally charged throughout the domain:
\begin{equation}
    \sum_\pi z_\pi C_\pi = 0
    \label{eq:electroneutrality}
\end{equation}
These equations assume negligible interactions between the ion species outside of chemical reactions \cite{Sarkar2011}.\\ 

The $\mathrm{H}^+$ and $\mathrm{OH}^-$ ion concentrations are related through the water auto-ionization process:
\begin{equation}
    \mathrm{H}_2\mathrm{O} \xrightleftharpoons[k_{w}']{k_{w}} \mathrm{H}^+ + \mathrm{OH}^-
\end{equation}
which is implemented through the reaction term $R_\pi$ as:
\begin{equation}
    {R_{\mathrm{H}^+}}_1=R_{\mathrm{OH}^-} = k_{w}C_{\mathrm{H}_2\mathrm{O}} - k_{w}'C_{\mathrm{H}^+}C_{\mathrm{OH}^-}  = k_{eq} \left(K_w-C_{\mathrm{H}^+} C_{\mathrm{OH}^-} \right) \label{eq:water_react}
\end{equation}
where $K_w = 10^{-8} \; \mathrm{mol}^2/\mathrm{m}^6$ is the water auto-ionization constant and the variable $k_{eq}$ is given a sufficiently high value to enforce an equilibrium reaction; here, $k_{eq}=10^5\;\mathrm{m}^3/(\mathrm{mol}\cdot \mathrm{s})$. In addition, the $\mathrm{Fe}^{2+}$ ions react with water according to:
\begin{equation}
    \mathrm{Fe}^{2+} + \mathrm{H}_2\mathrm{O} \xrightleftharpoons[k_{fe}']{k_{fe}} \mathrm{FeOH}^+ + \mathrm{H}^+ \label{eq:Fe_H20}
\end{equation}
which in turn can react further through:
\begin{equation}
    \mathrm{FeOH}^{+} + \mathrm{H}_2\mathrm{O} \xrightharpoonup{k_{feoh}} \mathrm{Fe}(\mathrm{OH})_2 + \mathrm{H}^+ \label{eq:FeOH_H2O}
\end{equation}
with $\mathrm{Fe}(\mathrm{OH})_2$ assumed to not dissolve in water and its volume to be negligible compared to the domain size. These assumptions allow the concentration of $\mathrm{Fe}(\mathrm{OH})_2$ not to be explicitly simulated, and instead serve as a pathway for iron ions to exit the domain. We also assume that these solid reactants do not interfere with the surface reactions. Since Reactions \eqref{eq:Fe_H20} and \eqref{eq:FeOH_H2O} both produce $\mathrm{H}^+$, the pH of the electrolyte is expected to decrease in regions with large amounts of iron ion production due to corrosion. Each of the reactions are implemented through their associated reaction terms, given by:
\begin{align}
    R_{\mathrm{Fe}^{2+}}&=-k_{fe}C_{\mathrm{Fe}^{2+}}+k_{fe}'C_{\mathrm{FeOH}^-}C_{\mathrm{H}^+} \\
    R_{\mathrm{FeOH}^+}&=k_{fe}C_{\mathrm{Fe}^{2+}}-C_{\mathrm{FeOH}^-}(k_{feoh}+k_{fe}'C_{\mathrm{H}^+})\\
    {R_{\mathrm{H}^+}}_2&=k_{fe}C_{\mathrm{Fe}^{2+}}-C_{\mathrm{FeOH}^-}(k_{fe}'C_{\mathrm{H}^+}-k_{feoh}) \label{ref:RH2}
\end{align}
Here, one should note that the reaction term associated with $\mathrm{H}^+$ comprises both Eqs. \eqref{eq:water_react} and \eqref{ref:RH2}, such that $R_{\mathrm{H}^+} = {R_{\mathrm{H}^+}}_1+{R_{\mathrm{H}^+}}_2 $.\\

Finally, the electrolyte is subjected to the following boundary conditions on $\Gamma_{ext}$ and $\Gamma_{int}$:
\begin{align}
    &-D_{\pi}\left(\bm{\nabla}C_\pi+\frac{z_\pi F}{RT} C_\pi\bm{\nabla}\varphi\right) = \overline{J}_{\pi} \qquad \text{or} \qquad C_\pi = \overline{C}_\pi \\
    &\varphi = \overline{\varphi} 
\end{align}
with the externally imposed fluxes, concentrations, and electric fields given by $\overline{J}_{\pi}$, $\overline{C}_{\pi}$, and $\overline{\varphi}$, respectively.

\subsection{Interface interactions}
\begin{figure}
    \centering
    \includegraphics{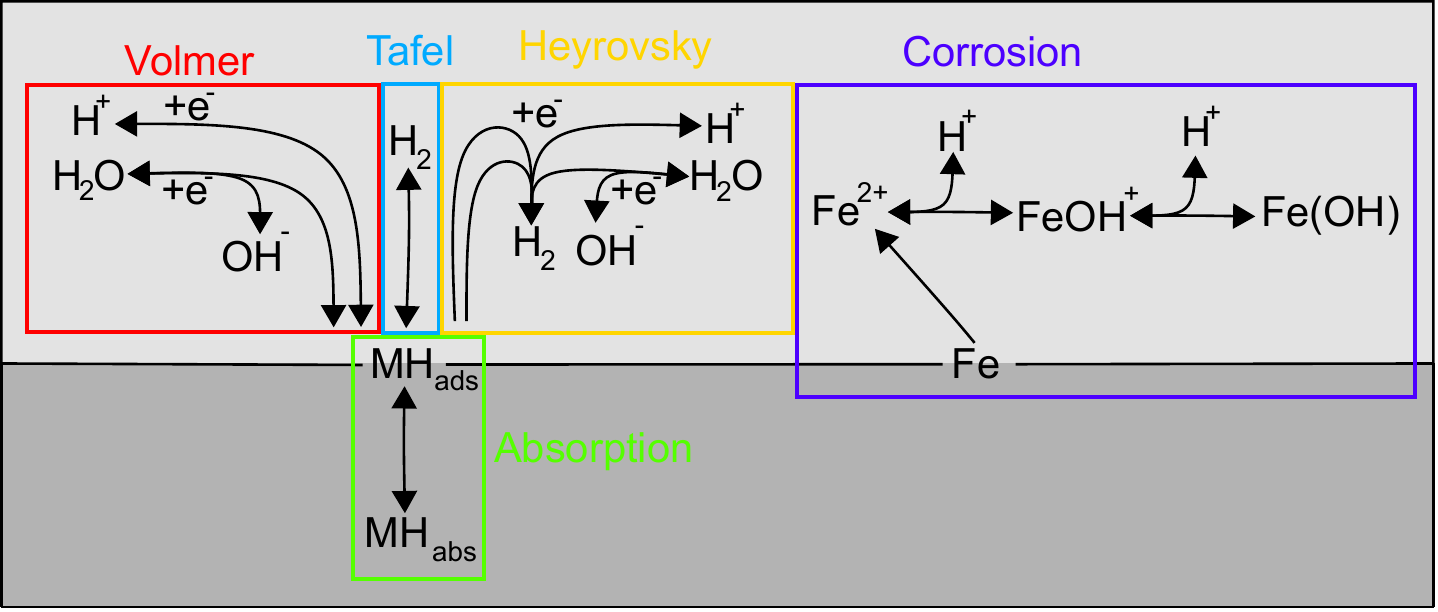}
    \caption{Schematic illustration of the hydrogen evolution (left) and corrosion (right) reactions occurring near the metal surface.}
    \label{fig:reactions_overview_figure}
\end{figure}
At the interface between the metal and the electrolyte, (electro-)chemical reactions convert the hydrogen within the electrolyte into surface hydrogen, defined through the adsorbed hydrogen concentration at the surface $\theta_{ads}$. For acidic electrolytes, the dominant reactions are given by the Volmer, Heyrovsky, Tafel, and absorption reactions:
\begin{alignat}{2}
 \text{Volmer:} && \mathrm{H}^+ + \mathrm{M} + \mathrm{e}^- &\xrightleftharpoons[k_{Va}']{k_{Va}} \mathrm{MH}_{ads} \label{react:1} \\
  \text{Heyrovsky:} && \qquad \mathrm{H}^+ + \mathrm{e}^- + \mathrm{MH}_{ads}&\xrightleftharpoons[k_{Ha}']{k_{Ha}} \mathrm{M} + \mathrm{H}_2 \label{react:2} \\
   \text{Tafel:} && 2 \mathrm{MH}_{ads} &\xrightleftharpoons[k_T']{k_T} 2\mathrm{M} + \mathrm{H}_2 \label{react:3} \\
   \text{Absorption:} && \mathrm{MH}_{ads} &\xrightleftharpoons[k_A']{k_A} \mathrm{MH}_{abs}  \label{react:4} 
\end{alignat}
whereas in non-acidic environments the alkaline versions of the Volmer and Heyrovsky reactions become more relevant:
\begin{alignat}{2}
    \text{Volmer:} &&  \mathrm{H}_2\mathrm{O} + \mathrm{M} + \mathrm{e}^- &\xrightleftharpoons[k_{Vb}']{k_{Vb}} \mathrm{MH}_{ads} + \mathrm{OH}^- \label{react:5} \\
   \text{Heyrovsky:} && \qquad  \mathrm{H}_2\mathrm{O} + \mathrm{e}^- + \mathrm{MH}_{ads}&\xrightleftharpoons[k_{Hb}']{k_{Hb}} \mathrm{M} + \mathrm{H}_2 + \mathrm{OH}^- \label{react:6} 
\end{alignat}
in which $\mathrm{M}$ denotes a metal atom interacting with hydrogen at the surface, and $k$ and $k'$ indicate forward and backward reaction constants. The reactions are schematically shown in Fig. \ref{fig:reactions_overview_figure}. The reaction rates for these reactions are given by \citep{Elhamid2000,Danaee2011,Liu2014,CS2020}:\\
\makebox[18cm][c]{
\hspace{-2.5cm}
\begin{minipage}{19cm}
\begin{alignat}{4}
\nonumber && && & \qquad\mathrm{Forward} &&  \qquad\qquad \mathrm{Backward} \\
    \mathrm{Volmer (acid):} && \; && \nu_{Va} &= k_{Va} C_{\mathrm{H}^+}(1-\theta_{ads})\exp{\left(-\alpha_{Va} \frac{\eta F}{RT}\right)}\;\;
    && \nu_{Va}' = k_{Va}' \theta_{ads}\exp{\left((1-\alpha_{Va}) \frac{\eta F}{RT}\right)} \label{eq:react1}\\
    \mathrm{Heyrovsky (acid):} && && \nu_{Ha} &= k_{Ha} C_{\mathrm{H}^+}\theta_{ads}\exp{\left(-\alpha_{Ha} \frac{\eta F}{RT}\right)}\qquad
    && \nu_{Ha}' = k_{Ha}' (1-\theta_{ads}) p_{\mathrm{H}_2} \exp{\left((1-\alpha_{Ha}) \frac{\eta F}{RT}\right)} \label{eq:react2}\\
    \mathrm{Tafel:} && && \nu_T &= k_T\theta_{ads}^2\qquad
    && \nu_T' = k_T' (1-\theta_{ads})\sqrt{p_{\mathrm{H}_2}} \label{eq:react3}\\
    \mathrm{Absorption:} && && \nu_A &= k_A (N_L - C_L)\theta_{ads}\qquad
    && \nu_A' = k_A' C_L (1-\theta_{ads}) \label{eq:react4}\\
    \mathrm{Volmer (base):} && && \nu_{Vb} &= k_{Vb} (1-\theta_{ads})\exp{\left(-\alpha_{Vb} \frac{\eta F}{RT}\right)}\qquad
    && \nu_{Vb}' = k_{Vb}' C_{\mathrm{OH}^-} \theta_{ads}\exp{\left((1-\alpha_{Vb}) \frac{\eta F}{RT}\right)} \label{eq:react5}\\
    \mathrm{Heyrovsky (base):} && && \nu_{Hb} &= k_{Hb} \theta_{ads}\exp{\left(-\alpha_{Hb} \frac{\eta F}{RT}\right)}\qquad
    && \nu_{Hb}' = k_{Hb}' (1-\theta_{ads}) p_{\mathrm{H}_2} C_{\mathrm{OH}^-} \exp{\left((1-\alpha_{Hb}) \frac{\eta F}{RT}\right)}  \label{eq:react6}
\end{alignat} \\ \end{minipage}}
where $\alpha$ is used to denote the charge transfer coefficients and the partial pressure of $\mathrm{H}_2$ is assumed to be negligible ($p_{\mathrm{H}_2} \approx 0$), allowing the backwards reaction rates for Reactions \eqref{eq:react2}, \eqref{eq:react3}, and \eqref{eq:react6} to be neglected. The electric overpotential $\eta$ is given by
\begin{equation}
    \eta=E_m - \varphi - E_{eq,\mathrm{H}}
    \label{eq:overpotential}
\end{equation}
where $E_m$ is the electric potential of the metal and $E_{eq,\mathrm{H}}$ denotes the equilibrium potential.\\ 

Equilibrium between the inflow and outflow fluxes can be assumed, thereby eliminating the need to treat the surface coverage as an independent degree of freedom (see \cite{CS2020}). Instead, we here choose to solve for $\theta_{ads}$, which simplifies the formulations for the fluxes at the interface and allows the surface reactions to be in a state of non-equilibrium. Accordingly, the evolution of the hydrogen surface coverage is given by:
\begin{equation}
    N_{ads} \dot{\theta}_{ads} - (\nu_{Va}-\nu_{Va}') + (\nu_{Ha}-\nu_{Ha}') +2 (\nu_T-\nu_T') + (\nu_A-\nu_A') - (\nu_{Vb}-\nu_{Vb}') + (\nu_{Hb}-\nu_{Hb}') = 0
    \label{eq:massbalanceinterface}
\end{equation}
with $N_{ads}$ being the number of adsorption sites per metal surface area.\\

On the other hand, the reaction rate for the corrosion of the metal surface (right side of Fig. \ref{fig:reactions_overview_figure}) is given by:
\begin{equation}
    \nu_{Fe} = k_c \exp{\left((1-\alpha_c)\frac{\eta F}{RT}\right)}
\end{equation}
where $k_c$ is the corrosion rate constant and the overpotential is estimated using the equilibrium potential for the corrosion reaction $E_{eq,\mathrm{Fe}}$ in Eq. \eqref{eq:overpotential}, as opposed to $E_{eq,\mathrm{H}}$. Since the focus is on hydrogen uptake, we assume the corrosion rate to be small. This allows the effects of the corrosion and $\mathrm{Fe}^{2+}$ concentration on the local pH to be included, without the need to model changes in domain boundaries due to metal dissolution. \\ 

The interactions with the electrolyte are included through the $\mathrm{H}^+$, $\mathrm{OH}^-$, and $\mathrm{Fe}^{2+}$ fluxes at the internal boundary $\Gamma_{int}$:
\begin{align}
    \overline{J}_{\mathrm{H}^+} &= -(\nu_{Va} - \nu_{Va}') - (\nu_{Ha} - \nu_{Ha}') \\
    \overline{J}_{\mathrm{OH}^-} &= \nu_{Vb} - \nu_{Vb}' + \nu_{Hb} - \nu_{Hb}'  \\
    \overline{J}_{\mathrm{Fe}^{2+}} &= \nu_{Fe}
\end{align}
and the interaction with the metal is accounted for through the absorbed hydrogen flux going into the metal:
\begin{equation}
    \overline{J} = \nu_A - \nu_A'
    \label{eq:JfluxAbs}
\end{equation}
These ion fluxes couple the $C_{\mathrm{H}^+}$, $C_{\mathrm{OH}^-}$, and $C_L$ concentrations at the internal interface $\Gamma_{int}$. Since the electrolyte pressure is negligible, $\Gamma_{int}$ is traction-free ($\overline{\mathbf{t}}=0$). Furthermore, the internal interface does not allow other ionic species to enter the metal ($\overline{J}_{\pi}=0$ for $\pi \neq \mathrm{H}^+,\mathrm{OH}^-,\mathrm{Fe}^{2+}$), and the metal has a constant and uniform electric potential due to its conductivity. Finally, the displacements in the material are assumed to be small, thereby neglecting changes in the size and shape of the electrolyte domain and preventing the need for deforming the mesh or re-meshing.\\

The interface reaction equations presented enable capturing the uptake of hydrogen and all relevant surface phenomena as a function of the local pH and overpotential. When combined with the equations describing the coupled deformation-diffusion behaviour of the metal (Section \ref{Sec:MetalDomain}) and the electrochemical behaviour of the electrolyte (Section \ref{Sec:ElectrolyteDomain}), hydrogen ingress can be quantified as a function of bulk environmental conditions (pH and potential difference). However, one should note that an important set of inputs to the model is the (backward and forward) reaction rate constants $k$ and $k'$. These have to be determined experimentally and vary from one material to another. Table \ref{tab:reactions_lit} reports reaction rate constants measured in the literature for pure Fe and Fe-based materials. The large scatter observed in the values reported for some reaction rate constants suggests a high sensitivity to the material and surface conditions, motivating the need for careful experimental measurements to improve the accuracy of modelling predictions. \enlargethispage{4\baselineskip}

\begin{table}
\begin{adjustwidth}{-1.5cm}{}
    \centering
    \caption{Forward and backward reaction rate constants reported or used in literature for pure Fe or Fe-based materials.}
    \label{tab:reactions_lit}
\begin{tabular}{ |l|l l|l l|l|  }
 \hline
  Reaction & Forward reaction rate constant $k$ & & Backward reaction rate constant $k'$ & \\
  \hline 
 $\nu_{Va}$ & \makecell[l]{$5\cdot10^{-10}$ \citep{Elhamid2000}; $5$ \citep{Turnbull1996, CS2020b}; $3.2\cdot10^{-5}$ \citep{Pickering1988z};\\ $2\cdot10^{-10}$, $1\cdot10^{-9}$, $2\cdot10^{-9}$, $2\cdot10^{-3}$\citep{Iyer1989}} & $ \mathrm{m}/\mathrm{s}$ & $0$ \citep{Turnbull1996} & $\mathrm{mol/(m}^2\mathrm{s)}$  \\
 \Xhline{0.1pt}  
 $\nu_{Ha}$ & $5\cdot10^4$ \citep{Turnbull1996,CS2020b}\tablefootnote{One should note that this magnitude appears to be inconsistent with other values listed in Table \ref{tab:reactions_lit}. For instance, causing the acidic Heyrovsky reaction to be dominant even in highly alkaline environments (see the analysis in Section \ref{sec:dimensionalAnalysis}), and resulting in virtually no hydrogen absorption within the metal.} & $\mathrm{m/s}\;\;$ & $0$ \citep{Turnbull1996} &$\mathrm{mol/(m}^2\mathrm{Pa\; s)}$  \\
 \Xhline{0.1pt}  
 $\nu_T$ & \makecell[l]{$7.8\cdot10^{-11}$ \citep{Bhardwaj2008}; $1.8\cdot10^{-3}$ \citep{Elhamid2000}; $22$\citep{Vecchi2018a};\\ $7\cdot10^{-7}$, $1\cdot10^{-3}$, $3\cdot10^{-2}$, $5\cdot10^{-2}$\citep{Iyer1989}} & $\mathrm{mol/(m}^2\mathrm{s)}$ & $0$ \citep{Turnbull1996} & $\mathrm{mol/(m}^2\mathrm{s \;Pa}^{1/2})$   \\
  \Xhline{0.1pt}  
 $\nu_A$ & \makecell[l]{$1.22\cdot10^5$ \citep{Turnbull1996,CS2020b}; $2.4\cdot10^{-12}$\citep{Elhamid2000b};\\$3.3\cdot10^{-10}$, $5.8\cdot10^{-9}$, $6.6\cdot10^{-8}$, $6\cdot10^{9}$\citep{Vecchi2018a}} & $\mathrm{m/s}$ & $8.8\cdot10^9$\citep{Turnbull1996,CS2020b}; $1.9\cdot10^{-5}$\citep{Elhamid2000b} & $\mathrm{m/s}$ \\
  \Xhline{0.1pt}  
 $\nu_{Vb}$ & $10^{-4}$\citep{Hitz2002};$8.29\cdot10^{-8}$ \citep{Bhardwaj2008}; $6.5\cdot10^{-3}$ \citep{Pickering1988z} & $\mathrm{mol/(m}^2\mathrm{s})$ & $2.84\cdot10^{-10}$ \citep{Bhardwaj2008}; $\mathcal{O}(10^{-7})$ \citep{Hitz2002} & $\mathrm{m/s}$ \\
  \Xhline{0.1pt}  
 $\nu_{Hb}$ & $10^{-7}$ \citep{Hitz2002}; $1.9\cdot10^{-10}$ \citep{Bhardwaj2008}& $\mathrm{mol/(m}^2\mathrm{s)}$ & $0$ \citep{Bhardwaj2008}; $5.5\cdot10^{-12}$ \citep{Hitz2002} & $\mathrm{m/(Pa \;s)}$ \\
 \hline
\end{tabular}
\end{adjustwidth}
\end{table}

\subsubsection{Regimes of relevance of individual interface reactions}
\label{sec:dimensionalAnalysis}

\begin{figure}
    \centering
    \includegraphics[scale=1.1]{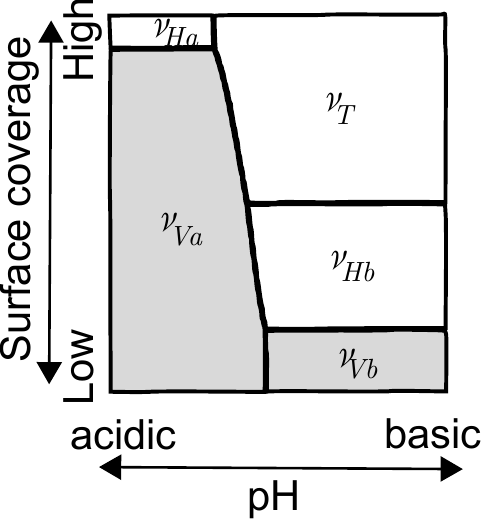}
    \caption{Schematic overview of the regimes of dominance of individual reactions as a function of the hydrogen surface coverage and the local pH. Reaction $\nu_A$ is not included. The regimes relevant to surface hydrogen producing reactions are shown in light grey, while white is used for regimes associated with surface hydrogen-consuming reactions.}
    \label{fig:RegimePlot}
\end{figure}

It should be noted that the generalised model presented results in significantly more reaction constants relative to other models which only include either the acidic or non-acidic hydrogen reactions \cite{Lee1971,Turnbull2015,Vecchi2018,Lasia2019}. Other simplifying assumptions for the hydrogen evolution reactions such as using a single rate-determining reaction \citep{Ma2020}, or assuming most backward reactions to be negligible \citep{Liu2014, Sun2019, Tang2020} are also often made to reduce the number of constants needed. In contrast, by including both acidic and non-acidic reactions within a single scheme, and not assuming a single rate-determining reaction step, our model is valid for the complete range of electrolyte pH and electric overpotentials. As a result, it can capture the large differences that occur between the open environment and occluded areas such as cracks. In addition, the generalised model presented encapsulates all other existing models, enabling its particularisation (e.g., to validate individual parts). However, not all reactions are relevant at the same time. A schematic illustration of the regimes of dominance of individual reactions is shown in Fig. \ref{fig:RegimePlot} as a function of the environment (pH, surface coverage). An estimate of the relative relevance of the acidic and base hydrogen producing Reactions, \eqref{eq:react1} and \eqref{eq:react5}, can be obtained by considering the ratio of their reaction rates:
\begin{equation}
   \frac{\nu_{Va}}{\nu_{Vb}} = \frac{C_{\mathrm{H}^+} k_{Va}}{k_{Vb}}\exp{\left((\alpha_{Vb}-\alpha_{Va})\frac{\eta F}{RT}\right)} \label{eq:scaling_15}
\end{equation}
Upon assuming $\alpha_{Vb}=\alpha_{Va}$, Eq. (\ref{eq:scaling_15}) implies that the non-acidic reaction becomes more important than the acidic one for $\mathrm{pH} >-\log_{10} (k_{Vb}/(1000k_{Va}))$ \footnote{The factor $1000$ is introduced due to the units of concentration used, $\mathrm{mol}/\mathrm{m}^3$, while the pH definition uses $\mathrm{mol}/\mathrm{L}$}. Below this pH, Reaction \eqref{eq:react1} will produce significantly more adsorbed hydrogen, whereas above this pH, Reaction \eqref{eq:react5} will determine the adsorbed hydrogen amount. Similar results are obtained for the $\mathrm{H}^2$ production through the Heyrovsky reactions \eqref{eq:react2} and \eqref{eq:react6}, with the non-acidic version becoming dominant for $\mathrm{pH} >-\log_{10} (k_{Hb}/(1000k_{Ha}))$.\\

An estimate for the total $\mathrm{H}^+$ diffusion within the electrolyte inside of a crack is given by $\nu_e = h D_{\mathrm{H}^+}C_{\mathrm{H}^+}^*/L$, with $C_{H^+}^*$ being an estimate for the $H^+$ concentration at the crack mouth, and $h$ and $L$ respectively denoting the height and length of the crack \footnote{Here, 1D $\mathrm{H}^+$ transport is assumed, due to a concentration gradient $C_{\mathrm{H}^+}^*/L$ through a channel with height $h$, such that the total flux through this channel is $J = hD C_{\mathrm{H}^+}^*/L$}. This parameter can be used to estimate whether the acidic surface reaction is the rate-limiting step, or if instead $\mathrm{H}^+$ diffusion within the crack is what limits the amount of adsorbed hydrogen. The ratio between the two is given by:
\begin{equation}
    \frac{\nu_{Va}}{\nu_{e}} = \frac{k_{Va} L^2 \exp{(-\alpha_{Va}\frac{\eta F}{RT})}}{hD_{\mathrm{H}^+}} \label{eq:scaling_LH}
\end{equation}
with the surface reactions being the rate-limiting step when this ratio is much smaller than one. Hence, longer and sharper cracks translate into a smaller transport of $\mathrm{H}^+$ towards the crack tip. The dominance of $\mathrm{H}^+$ diffusion becomes more relevant for negative overpotentials, as these require more hydrogen ions to sustain the surface reactions. This also gives a first-order estimate on whether simulating the electrolyte is important. When the ratio $\nu_{Va}/\nu_{e}$ is close to 1 or higher, then the pH of the electrolyte changes significantly within the crack or pit. However, if $\nu_{Va}/\nu_{e} << 1$, a rather uniform distribution of $C_{H^+}$ is expected and resolving local changes in pH might not be necessary. It should also be noted that the non-acidic reaction is never limited by the transport of $\mathrm{H}^+$, and therefore should be mostly independent of the crack geometry.\\

Similarly, the diffusion of absorbed hydrogen within the metal can be described by $\nu_{m}=D_LC_L/L^*$, with $L^*$ being a characteristic length scale of the diffusion problem. By assuming the absorption reaction to occur fast, the surface coverage can be related to the lattice concentration through $C_L \approx \theta_{ads}k_A N_L/k_A'$. This allows the amount of adsorbed hydrogen that diffuses into the metal to be compared to the rates at which it combines into $\mathrm{H}^2$ as:
\begin{equation}
    \frac{\nu_{Ha}}{\nu_m} = \frac{k_{Ha} C_{\mathrm{H}^+} L^* \exp{(-\alpha_{Ha}\frac{\eta F}{RT})}}{D_L k_A N_L/k_A'} \qquad  \frac{\nu_T}{\nu_m} = \frac{k_T \theta_{ads} L^* }{D_L k_AN_L/k_A'}  \qquad \frac{\nu_{Hb}}{\nu_m} = \frac{k_{Hb} L^* \exp{(-\alpha_{Hb}\frac{\eta F}{RT})}}{D_L k_AN_L/k_A'}
\end{equation}
where values lower than one indicate that hydrogen entry is limited by the reaction rate, while values far above one would be indicative of a hydrogen ingress process being limited by diffusion within the metal, resulting in an increased surface coverage and hydrogen gas production on the surface. When the diffusive length scale $L^*$ is close to zero, such as on the onset of simulations as this scale can be estimated based on the elapsed time ($L^*=\sqrt{tD_L}$), large amounts of hydrogen diffuse into the metal. When the forward absorption reaction constant is much larger than the backwards constant, $k_A/k_A'$, a high lattice concentration will be obtained inside the metal, and large amounts of hydrogen will diffuse into the material. However, as $D_Lk_A/k_A'$ decreases, the $\mathrm{H}^2$ reactions start to become more dominant. For a low surface coverage, Reactions \eqref{eq:react2} and \eqref{eq:react6} will remove the adsorbed hydrogen, whereas for a higher surface coverage and low electric overpotential, Reaction \eqref{eq:react3} will be the dominant one. Since Reaction \eqref{eq:react3} is environment-independent (not a function of the pH or electrolyte potential), it does not require an accurate representation of the electrolyte. Thus, when the surface coverage is expected to be high, the effects of simulating the electrolyte compared to using simplifications are limited. However, when a lower pH electrolyte is present or large overpotentials (in absolute value) are expected, the electrochemical behaviour of the electrolyte needs to be explicitly simulated in order to accurately predict the amount of adsorbed hydrogen reacting towards $\mathrm{H}_2$ (instead of being absorbed into the metal).

\section{Weak forms and numerical implementation}
\label{sec:disc}
The governing equations are discretised using the finite element method, requiring these equations to be cast into their weak forms. For the metal domain, this is done by multiplying the momentum balance from Eq. \eqref{eq:mombalance} with the test function for the displacements, $\delta \mathbf{u}$, and the hydrogen mass balance from Eq. \eqref{eq:massbalance2} with the test function for the lattice hydrogen concentration, $\delta C_L$, and integrating both over $\Omega_m$, resulting in:
\begin{align}
    \int_{\Omega_m}& \bm{\nabla}^s \left(\delta \mathbf{u}\right) \mathcal{L}_{el}\bm{\nabla}^s\mathbf{u}\; \mathrm{d}\Omega_m - \int_{\Gamma_{ext}} \delta \mathbf{u} \; \overline{\mathbf{t}}\; \mathrm{d}\Gamma_{ext} = \mathbf{0} \label{eq:weak1}\\
    \begin{split}
    \int_{\Omega_m}& \delta C_L \left(1+ \sum_i \frac{N_T^i/N_L \; \; \exp{\left(E_b^i/RT\right)}}{\left(1+C_L/N_L \exp{\left(E_b^i/RT\right)}\right)^2} \right)\dot{C}_L + D_L \bm{\nabla} \left(\delta C_L\right)\bm{\nabla} C_L - \frac{D_L \overline{V}_H}{3RT}\bm{\nabla}\left(\delta C_L\right) C_L \text{tr}\left(\mathcal{L}_{el}\bm{\nabla}^s\mathbf{u}\right)\; \mathrm{d}\Omega_m \\
    &- \int_{\Gamma_{ext}} \delta C_L \overline{J}\; \mathrm{d}\Gamma_{ext} - \int_{\Gamma_{int}} \delta C_L \left( \nu_A-\nu_A'\right)\; \mathrm{d}\Gamma_{int} = 0
    \end{split} \label{eq:weak2}
\end{align}
where $\mathcal{L}_{el}$ is the linear-elastic stiffness matrix. Note that the boundary flux arising from the boundary condition (Eq. \eqref{eq:BCflux}) is divided into two parts: One associated with the exterior boundary $\Gamma_{ext}$, where a hydrogen flux $\overline{J}$ can be prescribed, and one in the internal boundary $\Gamma_{int}$, where the flux is due to the absorption reaction - see Eq. \eqref{eq:JfluxAbs}. This last term provides the coupling between the adsorbed hydrogen and the hydrogen in the metal lattice.\\

Similarly, the weak forms for the electrolyte, Eqs. \eqref{eq:nernstplanck}-\eqref{eq:electroneutrality}, are obtained by multiplying with the test function for the ion concentrations, $\delta C_\pi$, and the electric potential $\delta \varphi$. This results in the following weak forms for the $\mathrm{H}^+$ and $\mathrm{OH}^-$ mass balances:
\begin{align}
    \begin{split}
        \int_{\Omega_e} &\delta C_{\mathrm{H}^+} \dot{C}_{\mathrm{H}^+}+\delta C_{\mathrm{H}^+} \mathbf{v}^T\bm{\nabla}C_{\mathrm{H}^+}+D_{\mathrm{H}^+}\bm{\nabla}\left(\delta C_{\mathrm{H}^+}\right) \bm{\nabla}C_{\mathrm{H}^+} \\ &+ \frac{FD_{H^+}}{RT} \bm{\nabla}\left(\delta C_{\mathrm{H}^+}\right) C_{\mathrm{H}^+} \bm{\nabla} \varphi + k_{eq}\delta C_{\mathrm{H}^+}\left(K_w-C_{\mathrm{H}^+}C_{\mathrm{OH}^-}\right)\; \mathrm{d}\Omega_e  \\
        &-\int_{\Gamma_{ext}} \delta C_{\mathrm{H}^+} \overline{J}_{\mathrm{H}^+}\; \mathrm{d}\Gamma_{ext} - \int_{\Gamma_{int}} \delta C_{\mathrm{H}^+} \left(-(\nu_{Va}-\nu_{Va}')-(\nu_{Ha}-\nu_{Ha}')\right) \;\mathrm{d}\Gamma_{int}= 0 
    \end{split} \label{eq:weak3} \\
    \begin{split}
        \int_{\Omega_e} &\delta C_{\mathrm{OH}^-} \dot{C}_{\mathrm{OH}^-}+\delta C_{\mathrm{OH}^-} \mathbf{v}^T\bm{\nabla}C_{\mathrm{OH}^-}+D_{\mathrm{OH}^-}\bm{\nabla}\left(\delta C_{\mathrm{OH}^-}\right) \bm{\nabla}C_{\mathrm{OH}^-} + \frac{FD_{\mathrm{OH}^-}}{RT} \bm{\nabla}\left(\delta  C_{\mathrm{OH}^-}\right) C_{\mathrm{OH}^-} \bm{\nabla} \varphi \\ &+ k_{eq}\delta C_{\mathrm{OH}^-}\left(K_w-C_{\mathrm{H}^+}C_{\mathrm{OH}^-}\right)\; \mathrm{d}\Omega_e \\
        &-\int_{\Gamma_{ext}} \delta C_{\mathrm{OH}^-} \overline{J}_{\mathrm{OH}^-}\; \mathrm{d}\Gamma_{ext} - \int_{\Gamma_{int}} \delta C_{\mathrm{OH}^-} \left((\nu_{Vb}-\nu_{Vb}')+(\nu_{Hb}-\nu_{Hb}')\right) \;\mathrm{d}\Gamma_{int}= 0 \, ,
    \end{split} \label{eq:weak4}
\end{align}
the weak forms for the $\mathrm{Fe}^{2+}$ and $\mathrm{FeOH}^+$ mass balances:
\begin{align}
    \begin{split}
        \int_{\Omega_e} &\delta C_{\mathrm{Fe}^{2+}} \dot{C}_{\mathrm{Fe}^{2+}}+\delta C_{\mathrm{Fe}^{2+}} \mathbf{v}^T\bm{\nabla}C_{\mathrm{Fe}^{2+}}+D_{\mathrm{Fe}^{2+}}\bm{\nabla}\left(\delta C_{\mathrm{Fe}^{2+}}\right) \bm{\nabla}C_{\mathrm{Fe}^{2+}} + \frac{2FD_{\mathrm{Fe}^{2+}}}{RT} \bm{\nabla}\left(\delta C_{\mathrm{Fe}^{2+}}\right) C_{\mathrm{Fe}^{2+}} \bm{\nabla} \varphi \\
        & - k_{fe}\delta C_{\mathrm{Fe}^{2+}} C_{\mathrm{Fe}^{2+}} + k'_{fe} \delta C_{\mathrm{Fe}^{2+}} C_{\mathrm{FeOH}^+}C_{\mathrm{H}^+}\; \mathrm{d}\Omega_e -\int_{\Gamma_{ext}} \delta C_{\mathrm{Fe}^{2+}} \overline{J}_{\mathrm{Fe}^{2+}}\; \mathrm{d}\Gamma_{ext} - \int_{\Gamma_{int}} \delta C_{\mathrm{Fe}^{2+}}\nu_{Fe}  \;\mathrm{d}\Gamma_{int}= 0 
    \end{split} \\
    \begin{split}
        \int_{\Omega_e} &\delta C_{\mathrm{FeOH}^{+}} \dot{C}_{\mathrm{FeOH}^{+}}+\delta C_{\mathrm{FeOH}^{+}} \mathbf{v}^T\bm{\nabla}C_{\mathrm{FeOH}^{+}}+D_{\mathrm{FeOH}^{+}}\bm{\nabla}\left(\delta C_{\mathrm{FeOH}^{+}}\right) \bm{\nabla}C_{\mathrm{FeOH}^{+}} \\ &+ \frac{FD_{\mathrm{FeOH}^{+}}}{RT} \bm{\nabla}\left(\delta C_{\mathrm{FeOH}^{+}}\right) C_{\mathrm{FeOH}^{+}} \bm{\nabla} \varphi +k_{fe}\delta C_{\mathrm{FeOH}^{+}} C_{\mathrm{Fe}^{2+}} +  \delta C_{\mathrm{FeOH}^{+}} C_{\mathrm{FeOH}^+} \left( k'_{fe}C_{\mathrm{H}^+}+k_{feoh}\right)\; \mathrm{d}\Omega_e \\ & -\int_{\Gamma_{ext}} \delta C_{\mathrm{FeOH}^+} \overline{J}_{\mathrm{FeOH}^+}\; \mathrm{d}\Gamma_{ext} = 0 \, ,
    \end{split} 
\end{align}
and the weak form for the mass balances of the other $\pi$ ion phases:
\begin{equation}
    \int_{\Omega_e} \delta C_{\pi} \dot{C}_{\pi}+\delta C_{\pi} \mathbf{v}^T\bm{\nabla}C_{\pi}+D_{\pi}\bm{\nabla}\left(\delta C_{\pi}\right) \bm{\nabla}C_{\pi} + \frac{z_{\pi}FD_{\pi}}{RT} \bm{\nabla}\left(\delta C_{\pi}\right) C_{\pi} \bm{\nabla} \varphi \; \mathrm{d}\Omega_e -\int_{\Gamma_{ext}} \delta C_{\pi} \overline{J}_{\pi}\; \mathrm{d}\Gamma_{ext} = 0 
    \label{eq:weak5}
\end{equation}
The weak form for the electroneutrality condition is given by:
\begin{equation}
    \int_{\Omega_e} \delta \varphi \sum_{\pi} z_\pi C_\pi \; \mathrm{d}\Omega_e = 0
    \label{eq:weak6}
\end{equation}

Finally, the weak form of the mass balance at the internal interface is obtained by multiplying Eq. \eqref{eq:massbalanceinterface} with $\delta \theta$:
\begin{equation}
\begin{split}
        \int_{\Gamma_{int}} N_{ads} \delta \theta \; \dot{\theta}_{ads} + \delta \theta \Big(- (\nu_{Va}-\nu_{Va}') + (\nu_{Ha}-\nu_{Ha}') +2 (\nu_T-\nu_T') \\ + (\nu_A-\nu_A') - (\nu_{Vb}-\nu_{Vb}') + (\nu_{Hb}-\nu_{Hb}')\Big) \; \mathrm{d}\Gamma_{int} = 0
        \label{eq:weak7}
        \end{split}
\end{equation}
This last weak form couples the metal domain to the electrolyte domain through its reaction rates. The electrolyte potential and concentrations, together with the surface coverage, determine the reaction rates $\nu_{Va}$, $\nu_{Ha}$, $\nu_{Vb}$, $\nu_{Hb}$, and $\nu_{T}$. In turn, the surface coverage and the lattice hydrogen concentration provide reaction rate $\nu_{A}$, which couples the metal and electrolyte domains. These weak forms are discretised using quadratic quadrilateral elements for all variables in both domains, except for the hydrogen surface coverage which is discretised using quadratic line elements. The implementation is performed in the commercial finite element package \texttt{COMSOL Multiphysics}. The built-in tertiary current module \citep{COMSOL2020, Dickinson2014} is used for the electrolyte, while a new interface is developed using the physics builder to implement the interface reactions and the hydrogen transport inside the metal\footnote{The computational platform developed is made freely available at \url{www.empaneda.com/codes}}. The temporal discretisation was performed using a backward difference method. A mesh sensitivity study is conducted for all case studies so as to ensure reporting mesh-objective results. The number of DOFs employed ranged between 230,000 (Section \ref{sec:verif1}) and 390,000 (Section \ref{sec:results}). A full Newton-Raphson scheme was used to obtain converged solutions for the non-linear system.

\section{Verification case studies}
\label{sec:verif}
To verify the numerical implementation and physical behaviour of the model, we compare our results to two benchmark case studies: a numerical study simulating localised corrosion and its effect on the pH \cite{Sun2019}, and an experimental study measuring local pH within a large channel filled with metallic samples at set intervals \cite{Gangloff2014}.

\subsection{Numerical verification: localised corrosion}
\label{sec:verif1}
\begin{figure}
    \centering
    \includegraphics[width=7cm]{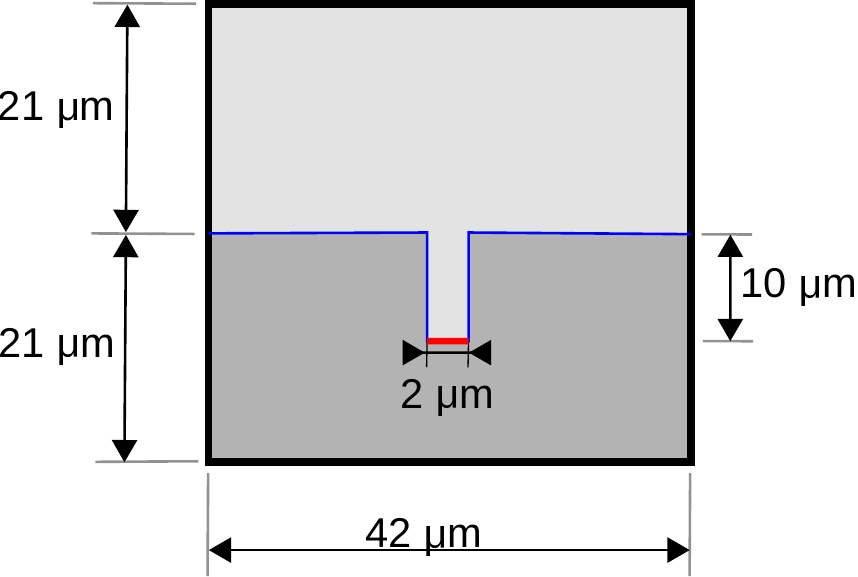}
    \caption{Localised corrosion verification case study. Geometry of the electrolyte (top) and metal (bottom) domains. Corrosion is allowed to occur at the red boundary (bottom of the notch), while the remaining metal-electrolyte boundaries (blue) are exposed to hydrogen reactions.}
    \label{fig:verif1}
\end{figure}

\begin{figure}
    \centering
    \begin{subfigure}{8cm}
         \centering
         \includegraphics[trim={0 0 0 0},clip]{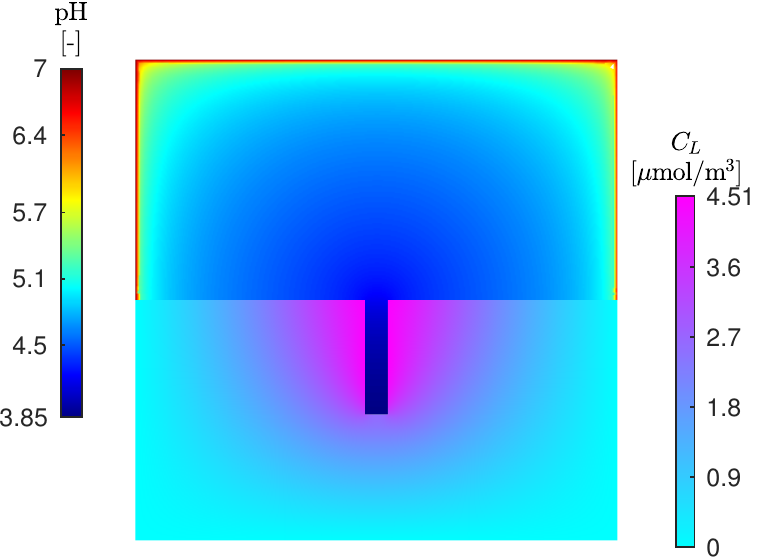}
         \caption{}
         \label{fig:verif2a}
    \end{subfigure}
    \begin{subfigure}{8cm}
         \centering
         \includegraphics[trim={0 0 0 0},clip]{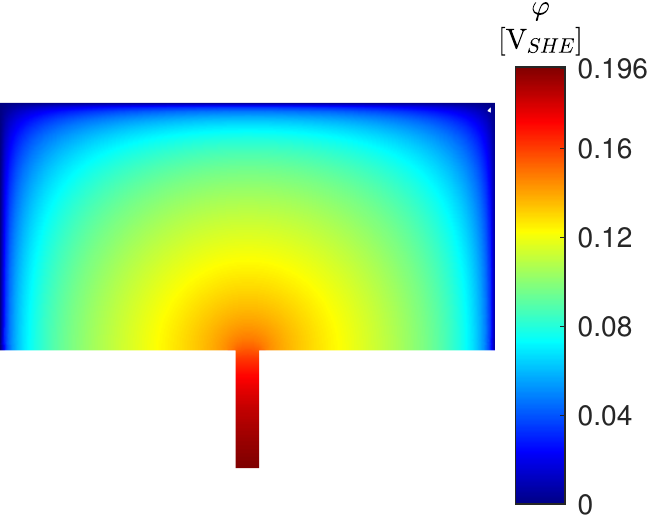}
         \caption{}
         \label{fig:verif2b}
     \end{subfigure}
    \caption{Localised corrosion verification case study. Predicted contours of (a) pH and lattice hydrogen concentration, and (b) electrolyte electric potential.}
    \label{fig:verif2}
\end{figure}

\begin{table}
\centering
 \caption{Localised corrosion verification case study. Parameters used, based on the work by Sun and Duddu \cite{Sun2019}.}
\label{table:verif1}
\begin{tabular}{ |l l||l|  }
 \hline
  Parameter & & Value\\
 \hline
 $\mathrm{H}^+$ diffusion coefficient & $D_{\mathrm{H}^+}$ & $9.3\cdot10^{-9}\;\mathrm{m}^2/\mathrm{s}$ \\
 $\mathrm{OH}^-$ diffusion coefficient & $D_{\mathrm{OH}^-}$ & $5.3\cdot10^{-9}\;\mathrm{m}^2/\mathrm{s}$ \\
 $\mathrm{Na}^+$ diffusion coefficient & $D_{\mathrm{Na}^+}$ & $10^{-9}\;\mathrm{m}^2/\mathrm{s}$ \\
 $\mathrm{Cl}^-$ diffusion coefficient & $D_{\mathrm{Cl}^-}$ & $10^{-9}\;\mathrm{m}^2/\mathrm{s}$ \\
$\mathrm{Fe}^{2+}$ diffusion coefficient & $D_{\mathrm{Fe}^{2+}}$ & $10^{-9}\;\mathrm{m}^2/\mathrm{s}$ \\
$\mathrm{FeOH}^+$ diffusion coefficient & $D_{\mathrm{FeOH}^+}$ & $10^{-9}\;\mathrm{m}^2/\mathrm{s}$ \\
\hline
Surface adsorption sites & $N_{ads}$ & $10^{-4}\;\mathrm{mol}/\mathrm{m}^2$ \\
Lattice sites & $N_L$ & $10^6\;\mathrm{mol}/\mathrm{m}^3$ \\
Lattice diffusion coefficient & $D_L$ & $10^{-9}\;\mathrm{m}^2/\mathrm{s}$ \\
Trap concentrations & $N_T$ & $[2.5,\; 1.0]\;\mathrm{mol}/\mathrm{m}^3$ \\
Binding energies & $E_b$ & $[15,\;30]\;\mathrm{kJ}/\mathrm{mol}$\\
 \hline
\end{tabular}
\end{table}
\begin{table}
\centering
 \caption{Localised corrosion verification case study. Reaction rates used, based on the work by Sun and Duddu \cite{Sun2019}.}
 \label{table:verif1b}
\begin{tabular}{ |l|l|l|l|l|  }
 \hline
  Reaction & $k$ & $k'$ & $\alpha$ & $E_{eq}$\\
 \hline
 $\nu_{Va}$ & $2.07\cdot10^{-12}\; \mathrm{m}/\mathrm{s}$ & $0\;\mathrm{mol/(m}^2\mathrm{s})$ & $0.5$ & $0\;\mathrm{V}_{SHE}$\\
 $\nu_{Ha}$ & $0 \;\mathrm{m/s}$ & $0 \;\mathrm{mol/(m}^2\mathrm{Pa \;s})$ & $0.5$ & $0\;\mathrm{V}_{SHE}$\\
 $\nu_T$ & $0 \;\mathrm{mol/(m}^2\mathrm{s})$ & $0 \;\mathrm{mol/(m}^2\mathrm{s \;Pa}^{1/2})$ & $-$ & $-$ \\
 $\nu_A$ & $1.2\cdot10^5 \;\mathrm{m/s}$ & $8.8\cdot10^9 \;\mathrm{m/s}$ & $-$ & $-$\\
 $\nu_{Vb}$ & $8.29\cdot10^{-15} \; \mathrm{mol/(m}^2\mathrm{s})$ & $0 \;\mathrm{m/s}$ & $0.5$ & $0\;\mathrm{V}_{SHE}$\\
 $\nu_{Hb}$ & $0 \;\mathrm{mol/(m}^2\mathrm{s})$ & $0 \;\mathrm{m/(Pa \;s)}$ & $0.5$ & $0\;\mathrm{V}_{SHE}$\\
 $\nu_{Fe}$ & $2.8\cdot10^6\;\mathrm{mol}/\mathrm{(m}^2\mathrm{s})$ & $-$ & $0$ & $0\;\mathrm{V}_{SHE}$\\
 \hline
 \hline
$k_{fe}/k_{fe}'$ &$1.625\cdot10^{-4}\;\mathrm{mol}/\mathrm{m}^3$ & &  & \\
$k_{feoh}$ & $0 \;\mathrm{s}^{-1}$ & & &\\
\hline
\end{tabular}
\end{table}

The geometry, parameters and initial and boundary conditions of the first verification case follow the computational study by Sun and Duddu \cite{Sun2019}. As shown in Figure \ref{fig:verif1}, the boundary value problem consists of a square domain containing a notched metallic sample. Corrosion is taking place at the bottom of the notch, while hydrogen reactions are allowed to occur at the remaining boundaries. The electrolyte consists of a solution of $\mathrm{NaCl}$ at an initial concentration of $C_{\mathrm{Na}^+}=C_{\mathrm{Cl}^-}=1\; \mathrm{mol}/\mathrm{m}^3$ and an initial pH of 7 ($C_{\mathrm{H}^+}=C_{\mathrm{OH}^-}=10^{-4}\; \mathrm{mol}/\mathrm{m}^3$), with these initial conditions also imposed as boundary conditions on the external boundaries. The initial and boundary concentrations of iron $\mathrm{Fe}^{2+}$ ions and $\mathrm{FeOH}^+$ reactants are zero. The parameters used for this simulation are given in Table \ref{table:verif1}, with the reaction constants being given in Table \ref{table:verif1b}. On the external boundaries, a constant electrolyte potential $\overline{\varphi}=0\;\mathrm{V}_{SHE}$ is imposed, the lattice hydrogen concentration is set to $\overline{C}_L=0\;\mathrm{mol}/\mathrm{m}^3$, and no mechanical load is applied. A constant electric potential of $E_m=-0.2\;\mathrm{V}_{SHE}$ is assigned to the metal. In contrast to the reference solution, we also model the absorbed hydrogen transport inside the metal using a diffusion coefficient $D_L = 1\cdot10^{-9}\;\mathrm{m}^2/\mathrm{s}$, a lattice site density $N_L = 10^6\;\mathrm{mol}/\mathrm{m}^3$, and considering two types of hydrogen traps, with densities $N_T=[2.5,\; 1.0]\;\mathrm{mol}/\mathrm{m}^3$ and binding energies $E_b = [15,\;30]\;\mathrm{kJ}/\mathrm{mol}$.\\ 

The results obtained after steady-state is reached (at $t=12\;\mathrm{s}$) are shown in Fig. \ref{fig:verif2}, and are in good agreement with the results reported by Sun and Duddu \cite{Sun2019}; minimum pH of 3.85 and maximum electrolyte potential of $\varphi=0.196\;\mathrm{V}_{SHE}$ versus reference results of pH=3.824 and $\varphi=0.1973\;\mathrm{V}_{SHE}$ \cite{Sun2019}. The pH inside the crack decreases due to corrosion creating $\mathrm{Fe}^{2+}$ ions, which then react to produce $\mathrm{H}^+$ and $\mathrm{FeOH}^+$ faster than the hydrogen reactions convert the $\mathrm{H}^+$ into absorbed hydrogen. While not explored in Ref. \cite{Sun2019}, our simulation shows that this pH drop and the applied boundary conditions lead to larger amounts of absorbed hydrogen in the notch, whereas the hydrogen reactions are slower on the boundaries far away from it. A similar effect is seen for the electric potential of the electrolyte, which increases locally due to the strong corrosion reaction taking place. This results in a reduction of the corrosion rate and in an acceleration of the hydrogen reactions. Since this effect is also most pronounced inside the notch, near the corroding surface, it also leads to a larger hydrogen uptake inside the notch compared to the exterior.  

\subsection{Experimental verification: local pH measurements in an artificial crevice cell}
\label{sec:verif2}
\begin{figure}
\centering
    \begin{subfigure}{15cm}
    \centering
    \includegraphics[width=12cm]{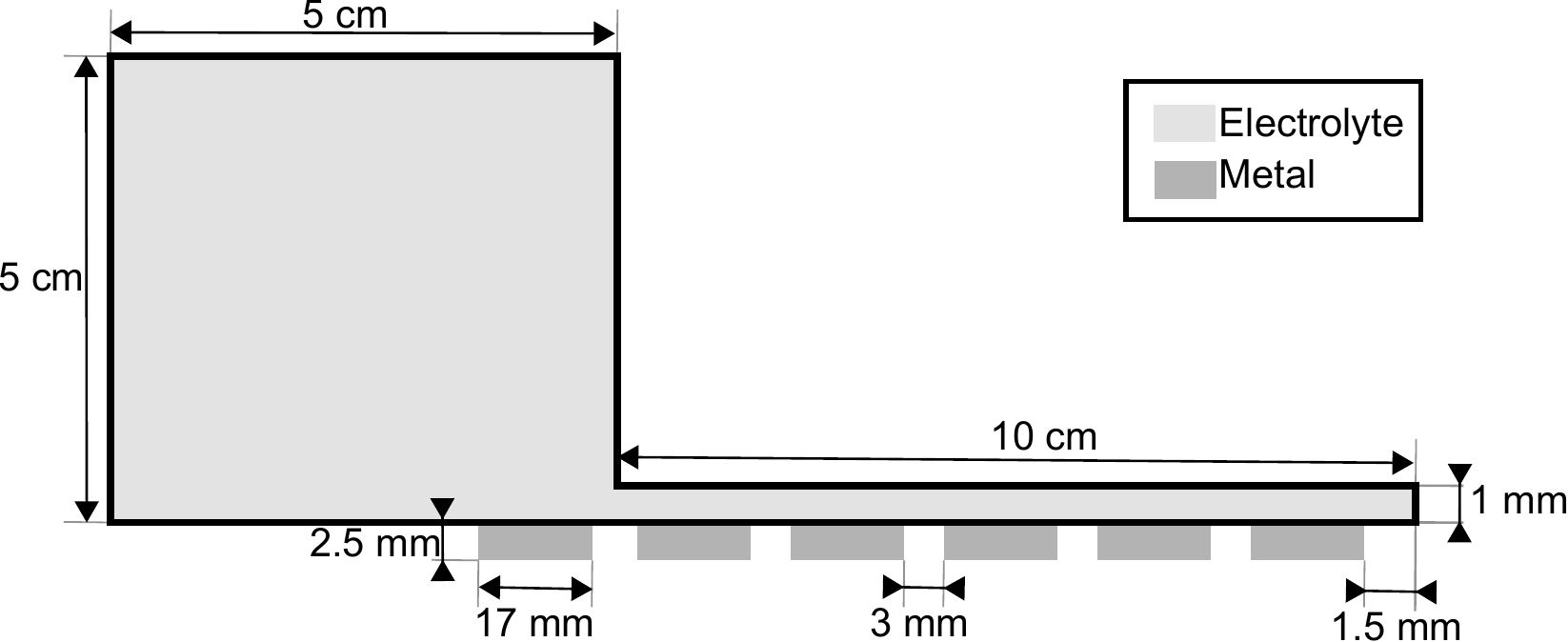}
    \subcaption{}
    \label{fig:verif3}
    \end{subfigure}
    \begin{subfigure}{15cm}
         \centering
         \includegraphics[width=14cm]{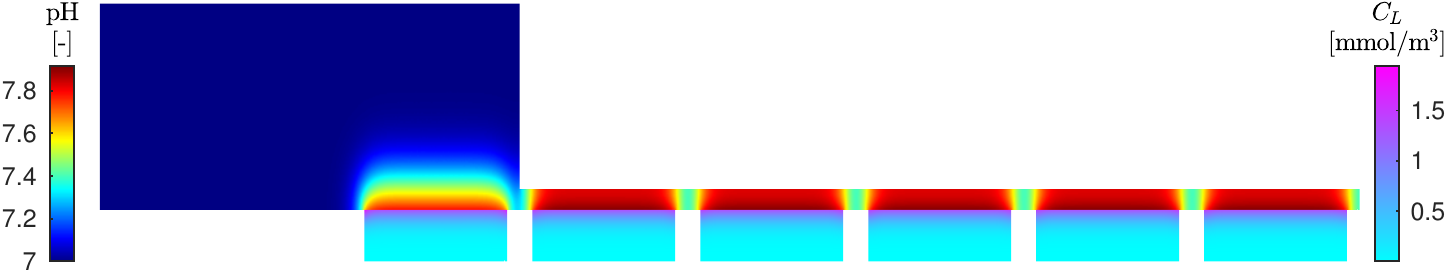}
         \caption{$E_m=-0.6\;\mathrm{V}_{SHE}$}
         \label{fig:verif4a}
    \end{subfigure}
    \begin{subfigure}{15cm}
         \centering
         \includegraphics[width=14cm]{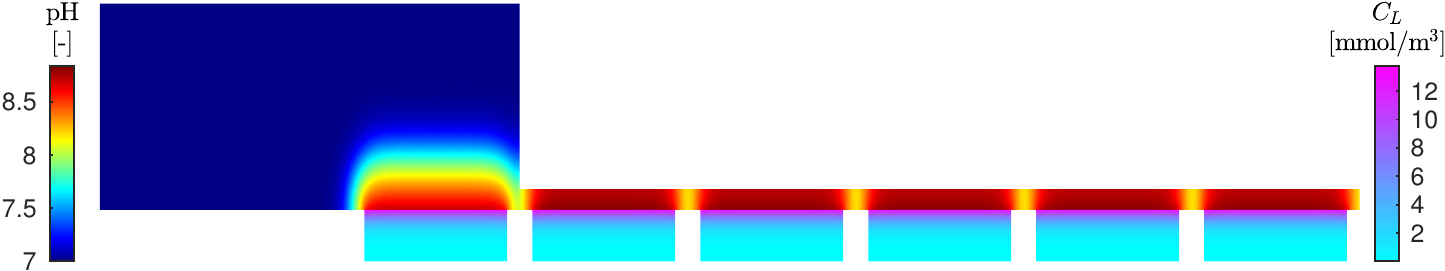}
         \caption{$E_m=-0.8\;\mathrm{V}_{SHE}$}
         \label{fig:verif4b}
     \end{subfigure}
     \begin{subfigure}{15cm}
         \centering
         \includegraphics[width=14cm]{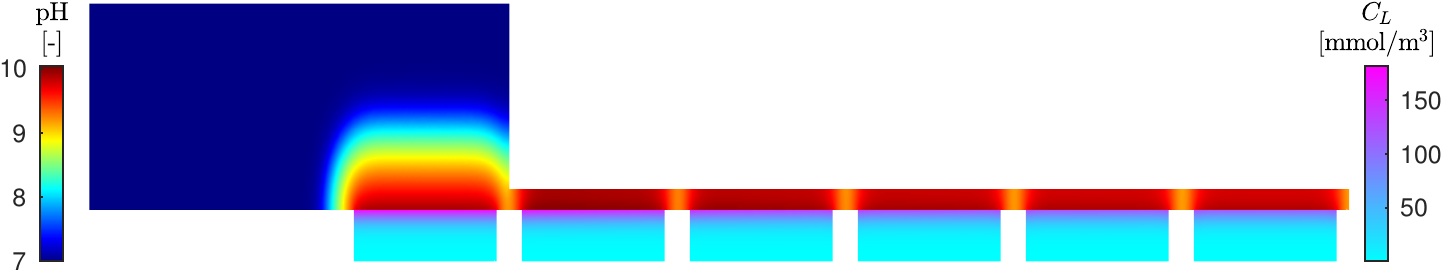}
         \caption{$E_m=-1\;\mathrm{V}_{SHE}$}
         \label{fig:verif4c}
     \end{subfigure}
    \caption{Experimental verification case study: (a) geometry of the artificial crevice cell used to measure local pH in Ref. \cite{Gangloff2014}, and predictions of pH and lattice hydrogen concentration at $t=300\;\mathrm{s}$ for (b) $E_m=-0.6\;\mathrm{V}_{SHE}$, (c) $E_m=-0.8\;\mathrm{V}_{SHE}$ and (d) $E_m=-1\;\mathrm{V}_{SHE}$.}
    \label{fig:verif4}
\end{figure}

The second verification case study aims at benchmarking model predictions with local pH measurements from the artificial crevice electrochemical cell developed by Gangloff \textit{et al.} \cite{Gangloff2014}. As shown in Fig. \ref{fig:verif3}, the testing configuration consists of an artificial opening of $100\times1 \; \mathrm{mm}$ attached to a large reservoir filled with the electrolyte. At the bottom of the opening, 6 metal samples with size $17\times2.5 \; \mathrm{mm}$ are present at regular intervals. An electric potential is applied to these metallic samples, and a neutral electric potential boundary condition is applied to the left and top ends of the reservoir. Since no reaction or diffusion constants are given for the metal used (Monel K-500, a Ni-based superalloy), we have estimated these by iterating the simulation results until a reasonable match was achieved for the case of $E_m=-1\;\mathrm{V}_{SHE}$, after which these parameters were used to predict the results for the $E_m=-0.6\;\mathrm{V}_{SHE}$ and $E_m=-0.8\;\mathrm{V}_{SHE}$ cases. The electrolyte consists of a $600\;\mathrm{mol}/\mathrm{m}^3$ $\mathrm{NaCl}$ solution at pH 7 ($C_{\mathrm{H}}^+=C_{\mathrm{OH}}^-=10^{-4}\;\mathrm{mol}/\mathrm{m}^3$). The diffusivity of the ion species in this electrolyte is $D_{\mathrm{H}^+}=9.3\cdot10^{-9}\;\mathrm{m}^2/\mathrm{s}$, $D_{\mathrm{OH}^-}=5.3\cdot10^{-9}\;\mathrm{m}^2/\mathrm{s}$ and $D_{\mathrm{Na}^+}=D_{\mathrm{Cl}^-}=10^{-9}\;\mathrm{m}^2/\mathrm{s}$. At the surface, the reactions are described through $k_{Va}=10^{-9}\;\mathrm{m}/\mathrm{s}$, $k_A=1.2\cdot10^{5}\;\mathrm{m}/\mathrm{s}$, $k_A'=8.8\cdot10^{8}\;\mathrm{m}/\mathrm{s}$, and $k_{Vb} = 10^{-15}\;\mathrm{mol}/(\mathrm{m}^2\mathrm{s})$, with all other reaction constants set to zero. Also, corrosion is neglected. Inside the metal, the lattice diffusion is given by $D_L=10^{-9}\;\mathrm{m}^2/\mathrm{s}$ and no trapping sites are considered.\\

The pH inside the opening and the lattice hydrogen concentration inside the metal are shown in Fig. \ref{fig:verif4}. Since the $E_m=-1\;\mathrm{V}_{SHE}$ simulation was used to calibrate the reaction constants, the pH inside the opening matches the pH measured in Ref. \cite{Gangloff2014} after $300\;\mathrm{s}$, $\mathrm{pH}\approx 10$. It is also observed that, at the time of the results, the hydrogen has barely diffused through the metal, and most of the $\mathrm{H}^+$ that reacted at the surface was already present at the onset of the simulation instead of being supplied though diffusion within the electrolyte. The simulation of the other two cases, using $E_m = -0.6\;\mathrm{V}_{SHE}$ and $E_m=-0.8\;\mathrm{V}_{SHE}$, leads to pH predictions of $\mathrm{pH}=7.7$ and $\mathrm{pH}=8.6$, respectively. These are in good agreement with the experimental measurements: pH values of 7 and 8, respectively. The small differences observed can be justified by the intrinsic experimental scatter; the values reported in \citep{Gangloff2014} for the initial conditions are ``$\mathrm{pH}=6\;\mathrm{to}\;7$", and it is therefore reasonable to assume that a similar error band is valid for the results within the crack. This verification exercise confirms that the model is capable of replicating experimental observations. 

\FloatBarrier
\section{Quantifying hydrogen ingress and metal-electrolyte interactions}
\label{sec:results}
\begin{figure}
    \centering
    \includegraphics[width=8cm]{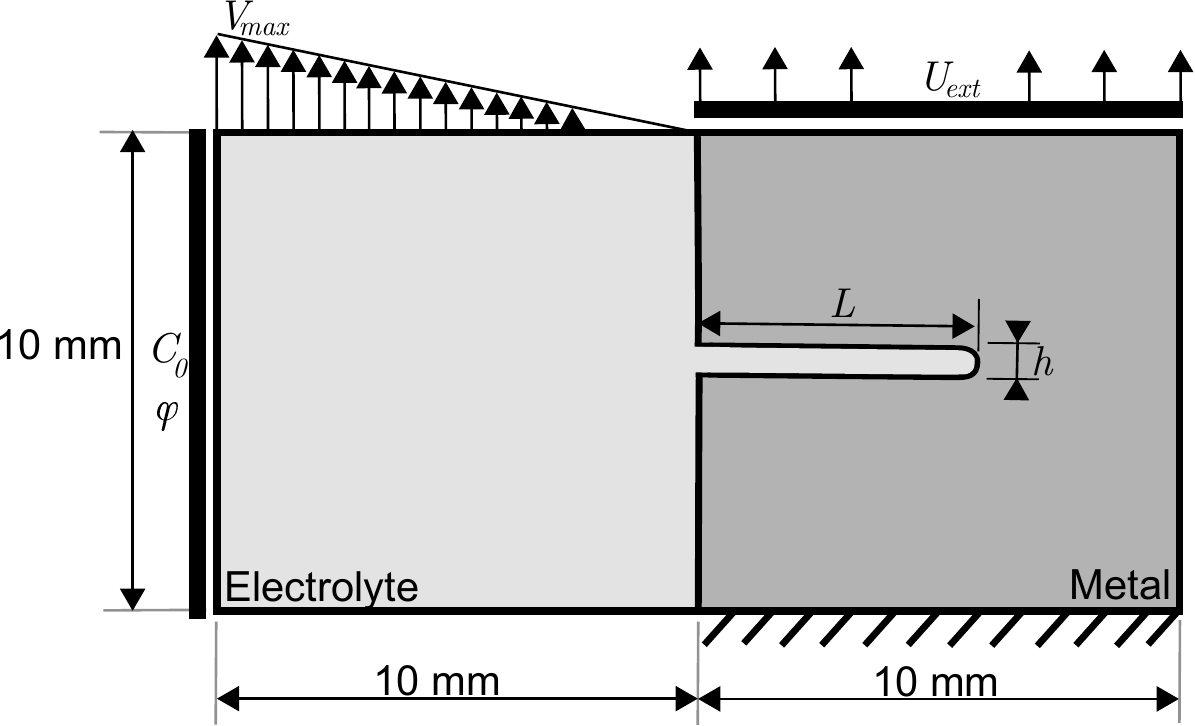}
    \caption{Overview of the used geometry and boundary conditions}
    \label{fig:geo}
\end{figure}

\begin{table}
 \caption{Material and ionic transport parameters used in the metal-electrolyte interaction studies of Section \ref{sec:results}.}
\label{table:params}
        \centering
\begin{tabular}{ |l l||l|  }
 \hline
  Parameter & & Value\\
 $\mathrm{H}^+$ diffusion coefficient & $D_{\mathrm{H}^+}$ & $9.3\cdot10^{-9}\;\mathrm{m}^2/\mathrm{s}$ \\
 $\mathrm{OH}^-$ diffusion coefficient & $D_{\mathrm{OH}^-}$ & $5.3\cdot10^{-9}\;\mathrm{m}^2/\mathrm{s}$ \\
 $\mathrm{Na}^+$ diffusion coefficient & $D_{\mathrm{Na}^+}$ & $1.3\cdot10^{-9}\;\mathrm{m}^2/\mathrm{s}$ \\
 $\mathrm{Cl}^-$ diffusion coefficient & $D_{\mathrm{Cl}^-}$ & $2\cdot10^{-9}\;\mathrm{m}^2/\mathrm{s}$ \\
$\mathrm{Fe}^{2+}$ diffusion coefficient & $D_{\mathrm{Fe}^{2+}}$ & $1.4\cdot10^{-9}\;\mathrm{m}^2/\mathrm{s}$ \\
$\mathrm{FeOH}^+$ diffusion coefficient & $D_{\mathrm{FeOH}^+}$ & $10^{-9}\;\mathrm{m}^2/\mathrm{s}$ \\
\hline
 Young's Modulus & $E$ & $200\;\mathrm{GPa}$\\
 Poisson ratio & $\nu$ & $0.25$\\
Partial molar volume & $\overline{V}_H$ & $2\cdot10^{-6}\;\mathrm{mol}/\mathrm{m}^3$\\
Surface adsorption sites & $N_{ads}$ & $10^{-4}\;\mathrm{mol}/\mathrm{m}^2$ \\
Lattice sites & $N_L$ & $10^6\;\mathrm{mol}/\mathrm{m}^3$ \\
Lattice diffusion coefficient & $D_L$ & $10^{-9}\;\mathrm{m}^2/\mathrm{s}$ \\
Trap concentrations & $N_T$ & $[2.5,\; 1.0]\;\mathrm{mol}/\mathrm{m}^3$ \\
Binding energies & $E_b$ & $[15,\;30]\;\mathrm{kJ}/\mathrm{mol}$\\
Temperature & $T$ & $293.15\;\mathrm{K}$\\
 \hline
\end{tabular}
\end{table}

\begin{table}
 \caption{Reaction rate constants used in the metal-electrolyte interaction studies of Section \ref{sec:results}.} \label{tab:reactionsused}
        \centering
\begin{tabular}{ |l|l|l|l|l|  }
 \hline
  Reaction & $k$ & $k'$ & $\alpha$ & $E_{eq}$\\
  \hline 
 $\nu_{Va}$ & $1\cdot10^{-4}\; \mathrm{m}/\mathrm{s}$ & $1\cdot10^{-10}\;\mathrm{mol/(m}^2\mathrm{s)}$ & $0.5$ & $0\;\mathrm{V}_{SHE}$ \\
 $\nu_{Ha}$ & $1\cdot10^{-10} \;\mathrm{m/s}\;\;$ & $0 \;\mathrm{mol/(m}^2\mathrm{Pa\; s)}$ & $0.3$ & $0\;\mathrm{V}_{SHE}$\\
 $\nu_T$ & $1\cdot10^{-6} \;\mathrm{mol/(m}^2\mathrm{s)}$ & $0 \;\mathrm{mol/(m}^2\mathrm{s \;Pa}^{1/2})$ & $-$ & $-$ \\
 $\nu_A$ & $1.2\cdot10^5 \;\mathrm{m/s}$ & $8.8\cdot10^9 \;\mathrm{m/s}$ & $-$ & $-$ \\ 
 $\nu_{Vb}$ & $1\cdot10^{-8} \; \mathrm{mol/(m}^2\mathrm{s})$ & $1\cdot10^{-13} \;\mathrm{m/s}$ & $0.5$ & $0\;\mathrm{V}_{SHE}$ \\
 $\nu_{Hb}$ & $8\cdot10^{-10} \;\mathrm{mol/(m}^2\mathrm{s)}$ & $0 \;\mathrm{m/(Pa \;s)}$ & $0.3$ & $0\;\mathrm{V}_{SHE}$ \\
 $\nu_{Fe}$ & $3.1\cdot10^{-10}\;\mathrm{mol}/\mathrm{(m}^2\mathrm{s)}$ & $-$ & $0$ & $-0.4\;\mathrm{V}_{SHE}$ \\
 \hline
$k_{fe}$ & $1\;\mathrm{s}$ & $10^{-3}\;\mathrm{m}^3/(\mathrm{mol}\;\mathrm{s})$ &  &  \\
$k_{feoh}$ & $10^{-2} \;\mathrm{s}^{-1}$ & & &  \\
$k_{eq}$ & $10^5\;\mathrm{m}^3/(\mathrm{mol}\;\mathrm{s})$ & & & \\
\hline
\end{tabular}
\end{table}

We use our electro-chemo-mechanical model to provide new insight by exploring the interactions between the metal and the electrolyte. The roles of the applied potential (Section \ref{sec:res_E}), fluid velocity (Section \ref{sec:fluid_V}) and defect geometry (Section \ref{sec:res_geo}) are investigated. To this end, we simulate two $10\times10\;\mathrm{mm}$ domains representing the electrolyte and the metal, with the metal containing an initial defect (pit, crack) of dimensions $L\times h$, as shown in Fig. \ref{fig:geo}. These defect dimensions are generally taken to be $L=5\;\mathrm{mm}$ and $h=0.4\;\mathrm{mm}$ but are also varied in Section \ref{sec:res_geo} to investigate their influence. The metallic sample is assumed to be uncharged with hydrogen at the onset of the simulations ($t=0$). In regards to its mechanical behaviour, the bottom of the sample is fixed, with both vertical and horizontal displacements constrained, and a vertical displacement of $U_{ext}=0.5\;\mu\mathrm{m}$ is applied at the top edge. The electrolyte has an initial $\mathrm{pH}=5$ with initial concentrations $C_{\mathrm{H}^+}=10^{-2}\;\mathrm{mol}/\mathrm{m}^3$, $C_{\mathrm{OH}^-}=10^{-6}\;\mathrm{mol}/\mathrm{m}^3$, $C_{\mathrm{Na}^+}=599.99\;\mathrm{mol}/\mathrm{m}^3$,  $C_{\mathrm{Cl}^-}=6\cdot10^{2}\;\mathrm{mol}/\mathrm{m}^3$, and $C_{\mathrm{Fe}^{2+}}=C_{\mathrm{FeOH}^+}=0\;\mathrm{mol}/\mathrm{m}^3$. Together with $\overline{\varphi}=0\;\mathrm{V}_{SHE}$, these concentrations are also prescribed on the left boundary as boundary conditions throughout the simulation. The electric potential of the metal is often kept at $E_m=0\;\mathrm{V}_{SHE}$ ($-0.24\;\mathrm{V}_{SCE}$) but also varied between $E_m=-0.7\;\mathrm{V}_{SHE}$ ($-0.94\;\mathrm{V}_{SCE}$) and $E_m=0.5\;\mathrm{V}_{SHE}$ ($0.26\;\mathrm{V}_{SCE}$). These values were chosen to span a wide range of environments, from positive potentials where corrosion reactions dominate, to negative potentials where hydrogen reactions govern the electrochemical behaviour. It should be noted that numerical convergence worsens significantly for applied potentials smaller than $-0.7\;\mathrm{V}_{SHE}$, due to the high reaction rates at the electrolyte-metal interface. The fluid velocity is generally assumed to be negligible (Section \ref{sec:res_E}) or assumed to change in a linear fashion from $V_{max}$ at the left edge of the electrolyte to zero at the electrolyte-metal interface (see Fig. \ref{fig:geo}). We investigate its influence by considering a value of $V_{max}=10\;\mathrm{mm}/\mathrm{s}$ in Section \ref{sec:res_geo} and by varying its magnitude from $V_{max}=0\;\mathrm{mm}/\mathrm{s}$ to $V_{max}=29\;\mathrm{mm}/\mathrm{s}$ in Section \ref{sec:fluid_V}. When a fluid velocity is included, additional boundary conditions are defined at the bottom and top of the electrolyte sub-domain: an inflow boundary condition at the bottom, setting the concentrations on this boundary equal to the initial conditions to emulate new electrolyte coming into the domain, and an outflow boundary condition at the top, restricting diffusion across this boundary while allowing advective species transport through it.\\

The material and ionic transport parameters used are given in Table \ref{table:params}, while the magnitudes of the reaction constants adopted are listed in Table \ref{tab:reactionsused}. Our choices aim at characterising the behaviour of Fe or Fe-based materials, for which sufficient data exists. In particular, our choices of reaction rate constants are based on those reported in the literature (see Table \ref{tab:reactions_lit}), focusing on the values reported for pure Fe and taking intermediate values within the range provided.\\


\subsection{Electric overpotential}
\label{sec:res_E}
\begin{figure}
    \centering
    \begin{subfigure}{12cm}
         \centering
         \includegraphics[width=12cm]{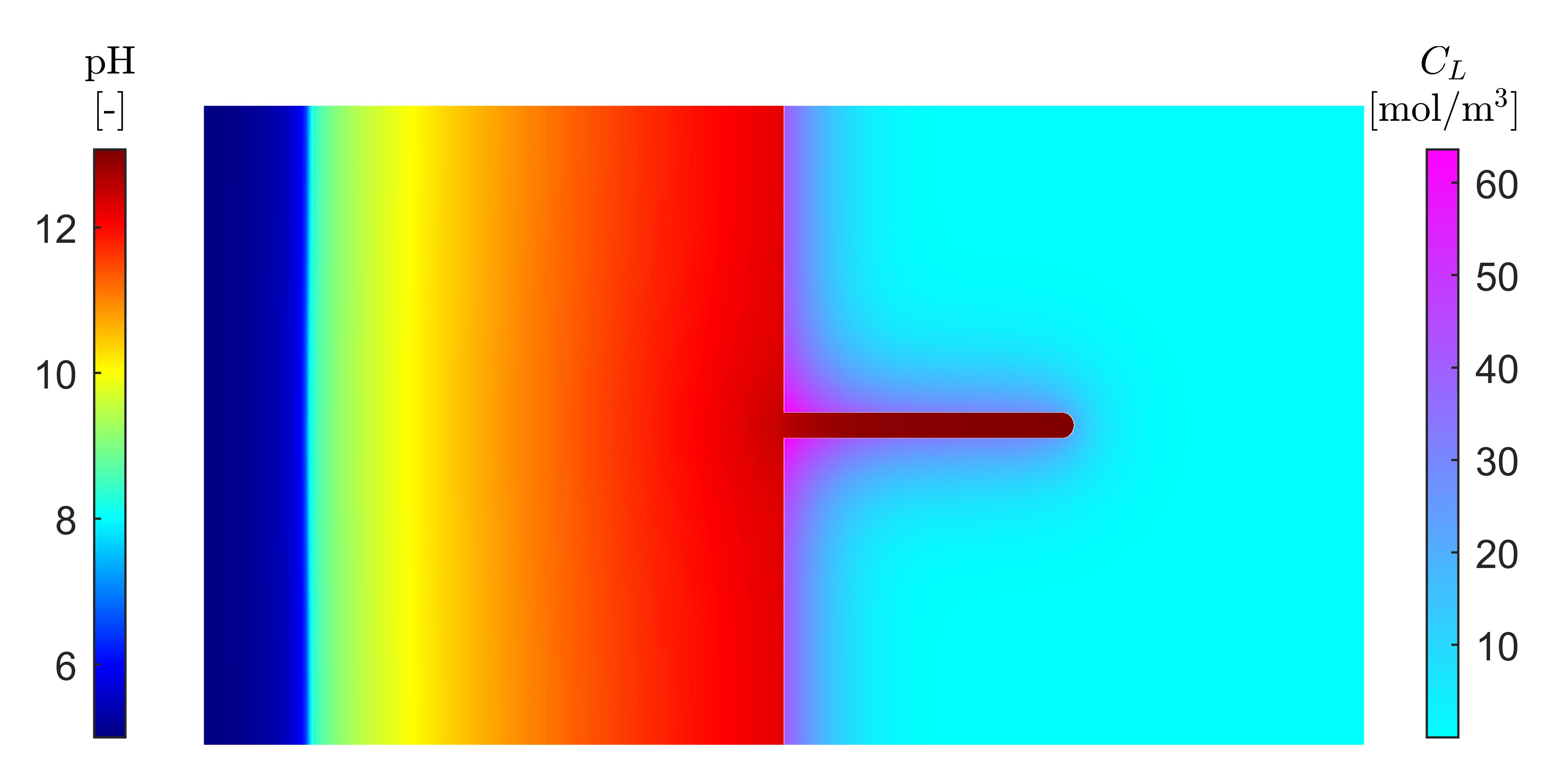}
         \caption{$E_m=-0.5\;\mathrm{V}_{SHE}$}
         \label{fig:Em_m05}
    \end{subfigure}
    \begin{subfigure}{12cm}
         \centering
         \includegraphics[width=12cm]{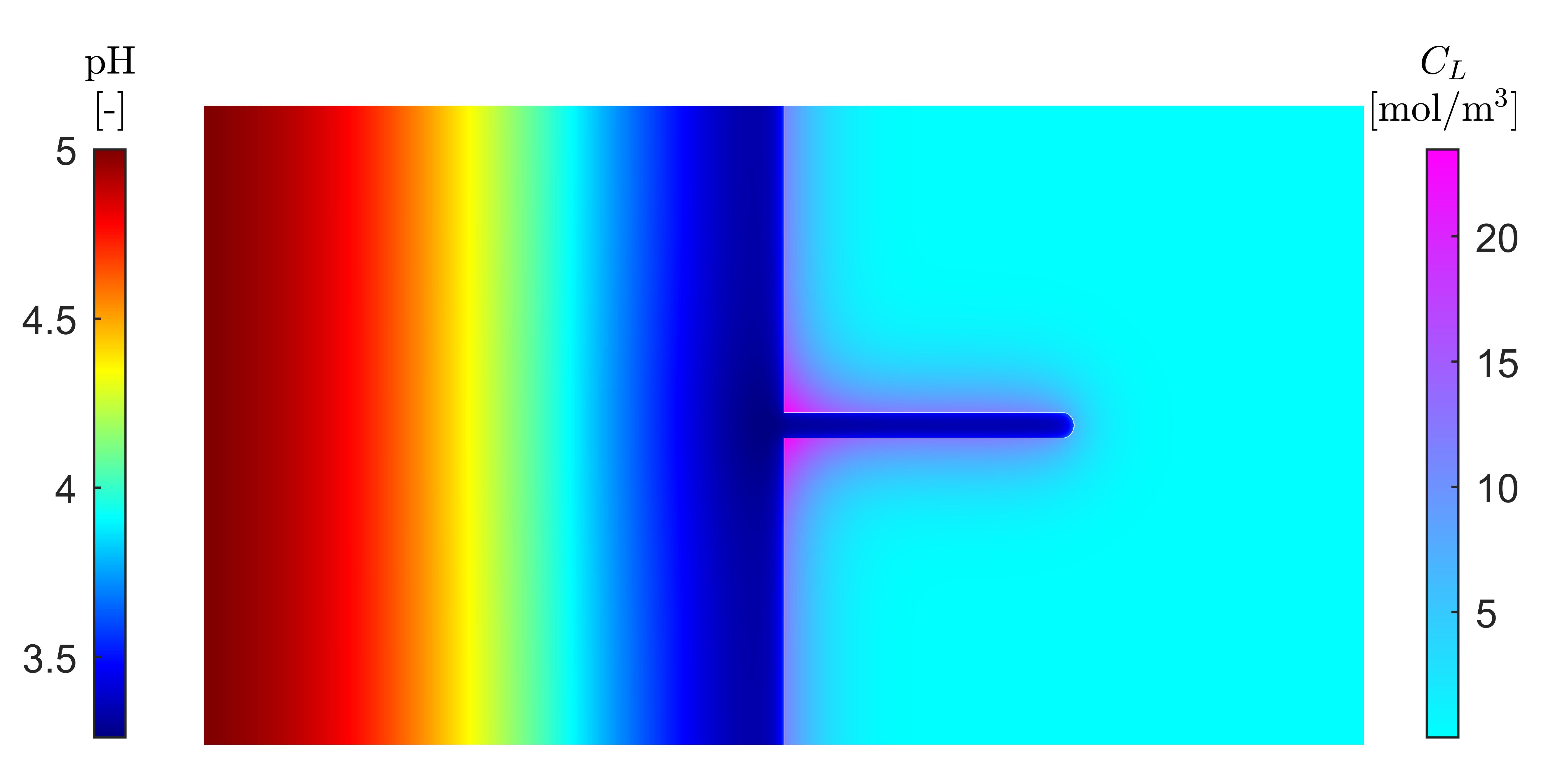}
         \caption{$E_m=0\;\mathrm{V}_{SHE}$}
         \label{fig:Em_0}
     \end{subfigure}
    \begin{subfigure}{12cm}
         \centering
         \includegraphics[width=12cm]{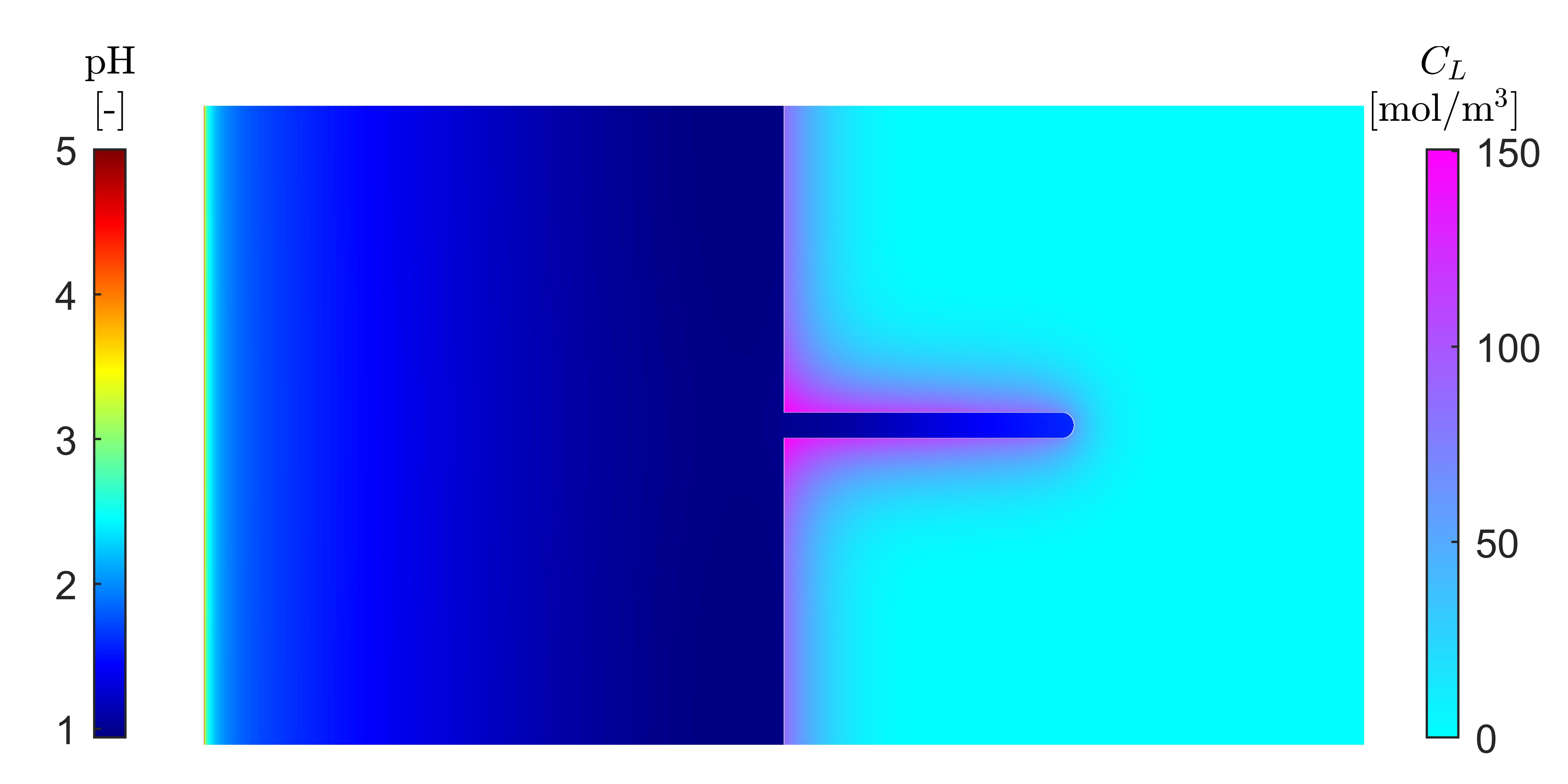}
         \caption{$E_m=0.5\;\mathrm{V}_{SHE}$}
         \label{fig:Em_p05}
     \end{subfigure}
    \caption{Metal-electrolyte interactions: influence of the applied potential $E_m$. Contours of pH (left, electrolyte domain) and lattice hydrogen concentration (right, metal domain) at a time $t=10\;\mathrm{min}$ for the following metal potentials: (a) $E_m=-0.5\;\mathrm{V}_{SHE}$, (b) $E_m=0\;\mathrm{V}_{SHE}$, and (c) $E_m=0.5\;\mathrm{V}_{SHE}$.}
    \label{fig:Em}
\end{figure}

\begin{figure}
    \centering
    \begin{subfigure}{8cm}
         \centering
         \includegraphics[width=8cm]{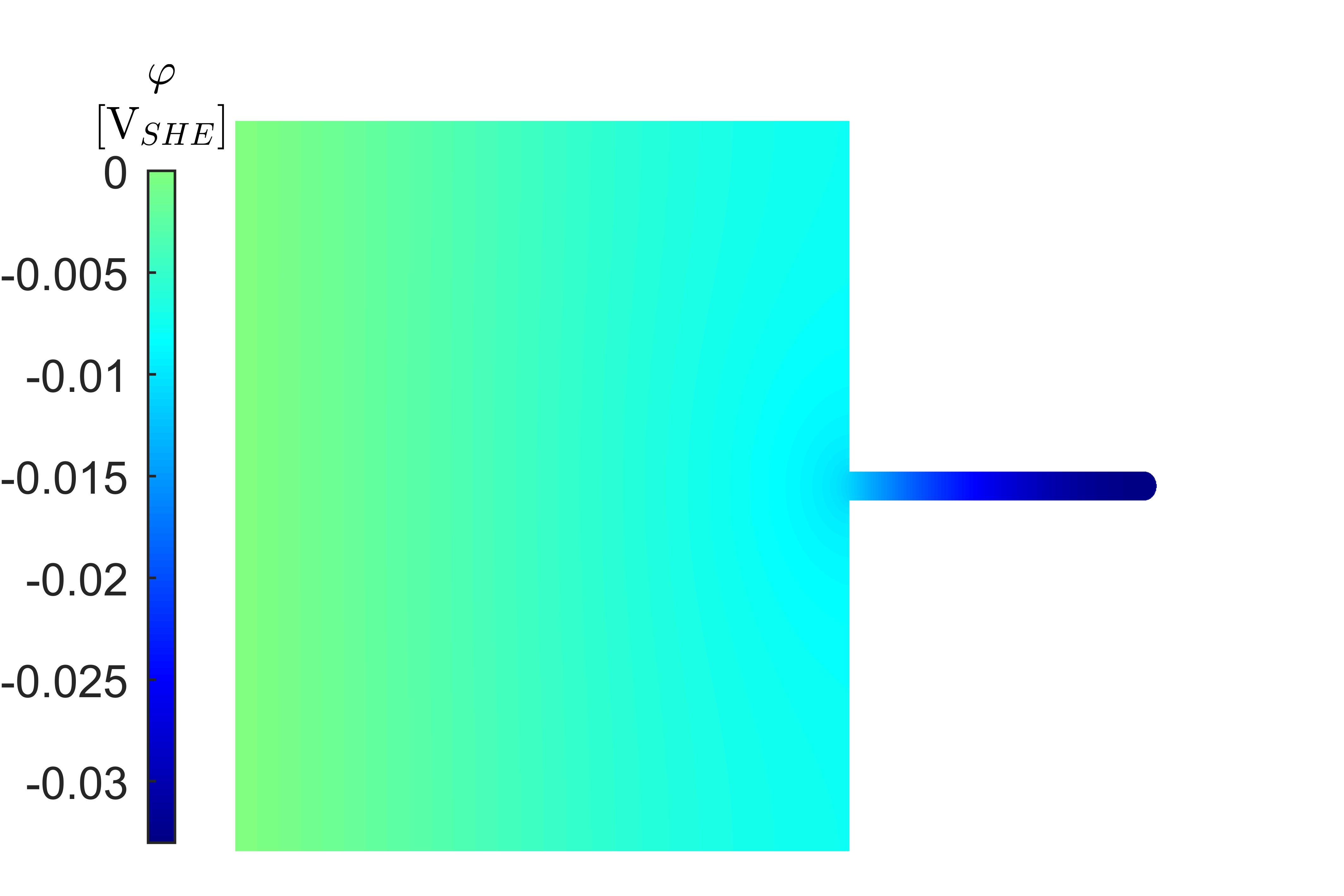}
         \caption{$E_m=-0.5\;\mathrm{V}_{SHE}$}
         \label{fig:Em_Phi_m05}
    \end{subfigure}
    \begin{subfigure}{8cm}
         \centering
         \includegraphics[width=8cm]{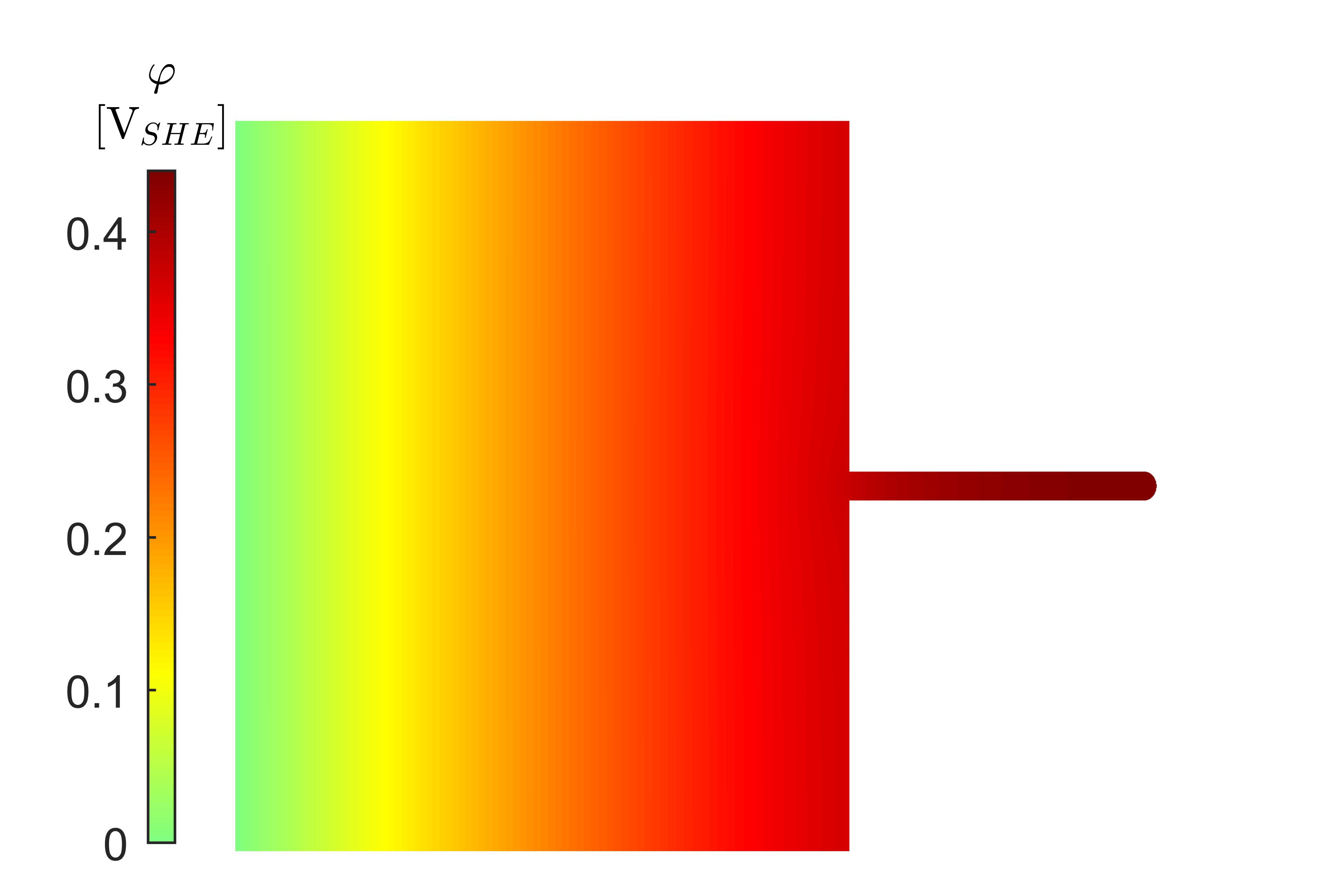}
         \caption{$E_m=0.5\;\mathrm{V}_{SHE}$}
         \label{fig:Em_Phi_0}
     \end{subfigure}
    \caption{Metal-electrolyte interactions: influence of the applied potential $E_m$. Contours of electrolyte potential $\varphi$ at a time $t=10\;\mathrm{min}$ for the following metal potentials: (a) $E_m=-0.5\;\mathrm{V}_{SHE}$ and (b) $E_m=0.5\;\mathrm{V}_{SHE}$.}
    \label{fig:Em_Phi}
\end{figure}

\begin{figure}
    \centering
    \includegraphics{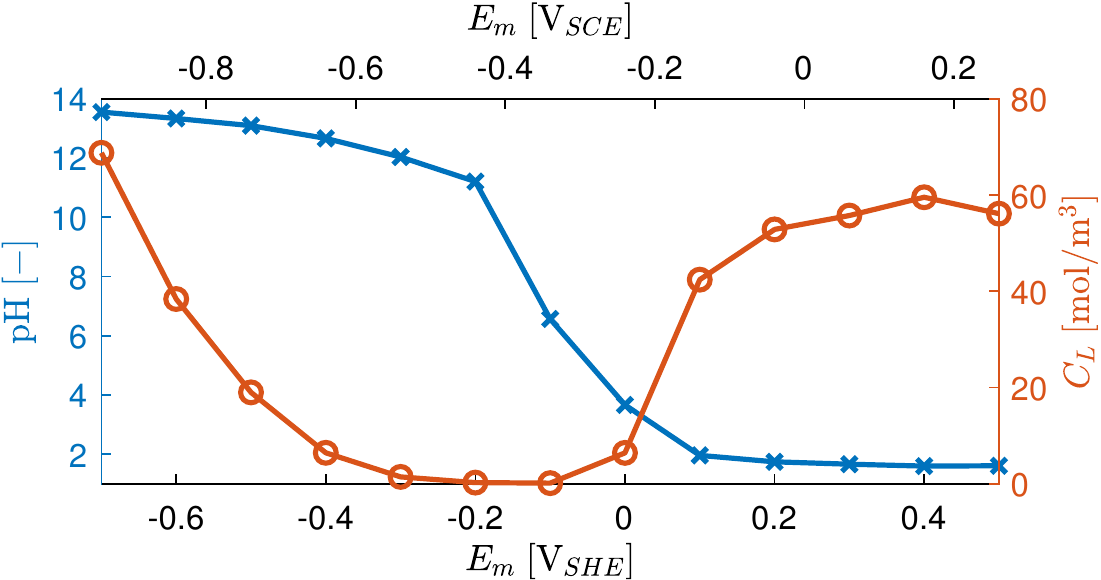}
    \caption{Metal-electrolyte interactions: Estimations of pH (blue crosses, left $y$-axis) and lattice hydrogen concentration (orange circles, right $y$-axis) at the crack tip after $t=10\;\mathrm{min}$, as a function of the applied potential.}
    \label{fig:Em_LatticeH}
\end{figure}

\begin{figure}
    \centering
    \includegraphics{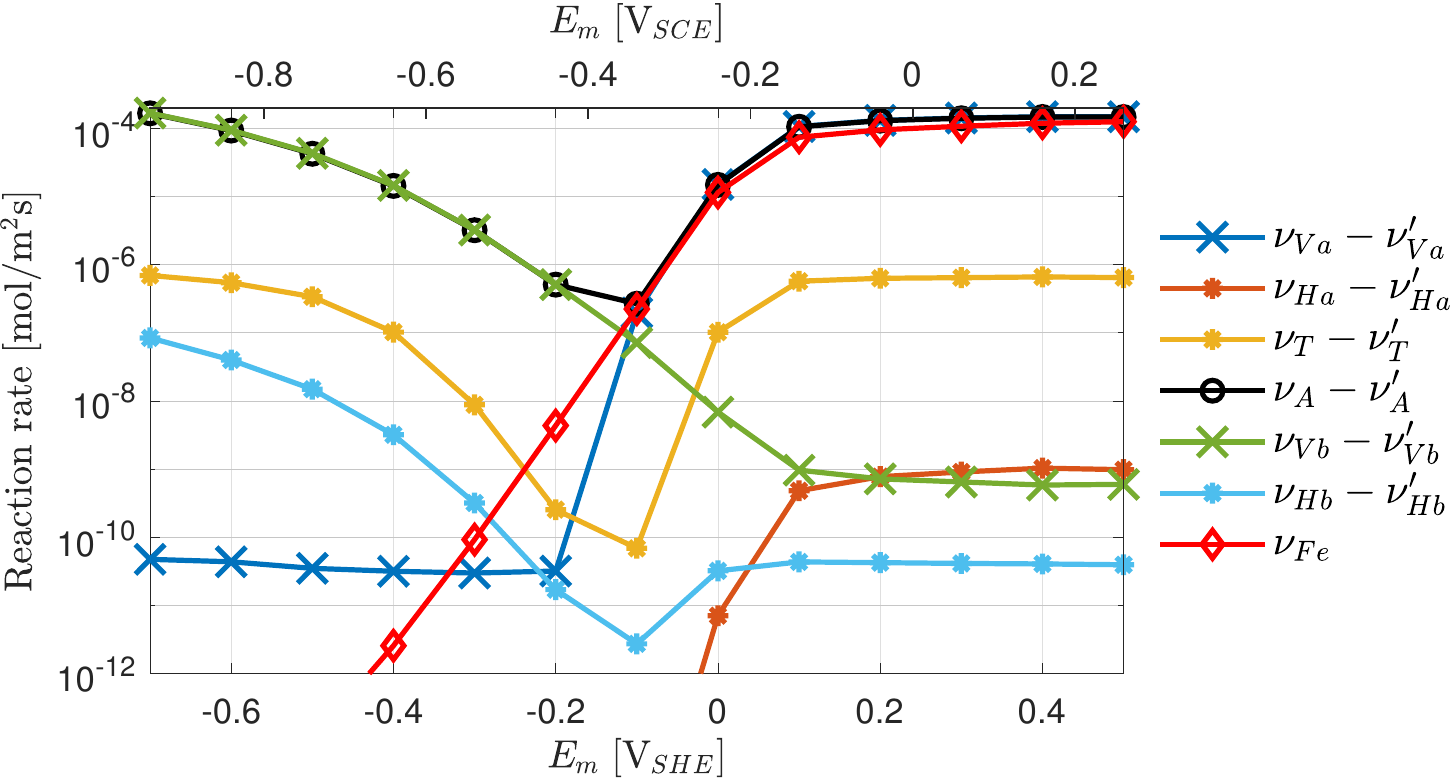}
    \caption{Metal-electrolyte interactions: Log-linear plot of the reaction rates at the crack tip, as a function of the applied potential after a time of $t=10\;\mathrm{min}$.}
    \label{fig:Em_ReactionRate}
\end{figure}

Corrosion reaction rates are reduced when a negative electric potential is applied to a metal, and thus this is one of the most commonly used methods to prevent corrosion (either by direct application, or through the addition of a sacrificial metal). However, these negative electric potentials increase the rate of the hydrogen reactions, augmenting the amount of absorbed hydrogen and the risk of experiencing hydrogen assisted failures \citep{Kehler2008,AM2016}. The results obtained are shown in Fig. \ref{fig:Em}. Specifically, contours of pH and lattice hydrogen are provided after $t=10$ min for selected values of the applied potential ($E_m$): $-0.5\;\mathrm{V}_{SHE}$, $0\;\mathrm{V}_{SHE}$, and $0.5\;\mathrm{V}_{SHE}$. The results show how the lowest applied potential considered accelerates the hydrogen reactions, up to the point where they absorb all available $\mathrm{H}^+$ ions, resulting in a high pH not just inside the pit but also on the exterior of the domain - see Fig. \ref{fig:Em}a. In addition, the non-acidic reaction in Eq. \eqref{eq:react5} is accelerated such that large amounts of hydrogen are absorbed inside the metal despite the high environmental pH, with this reaction producing additional $\mathrm{OH}^-$ ions to sustain the high pH of the electrolyte. In contrast, the corrosion reaction becomes relevant when a large positive electric potential is applied, producing iron ions which react within the electrolyte to produce additional $\mathrm{H}^+$, resulting in a lower pH near the metal (see Fig. \ref{fig:Em}c). This low pH strongly increases the adsorbed hydrogen produced through Reaction \eqref{eq:react1}, counteracting the reduction of hydrogen reaction rates associated with high electric potentials. Finally, when a neutral electric potential is applied, Fig. \ref{fig:Em}b, these two effects are balanced, with the pH being lowered by the corrosion reaction and raised by the hydrogen reactions. Another effect contributing to the differences is the geometry of the simulated domain. Near the entrance of the crack a higher lattice hydrogen concentration is observed since hydrogen is absorbed into the metal from both the exterior and crack faces. In contrast, at the crack tip the hydrogen concentration is slightly decreased due to the diffusion away from the crack tip causing the lattice hydrogen to spread over an increased area. \\

Changes in the applied potential also have an effect on the electrolyte potential, as shown in Fig. \ref{fig:Em_Phi}. For the case of a negative applied potential, $E_m=-0.5\;\mathrm{V}_{SHE}$, the corrosion reactions are non-existent and the hydrogen reactions are relatively slow (when compared with more aggressive cathodic potentials). As a result, small changes in the distribution of the electrolyte potential are observed, with the removal of positively charged $\mathrm{H}^+$ species translating into a small reduction in the electrolyte potential. In contrast, for the case of $E_m=0.5\;\mathrm{V}_{SHE}$ (Fig. \ref{fig:Em_Phi}b), the corrosion reaction dominates, causing large quantities of positively charged $\mathrm{Fe}^{2+}$ to enter the electrolyte at the interface. As a result, the electrolyte potential increases significantly, causing noticeable differences between the initial and boundary electric overpotential after just 10 minutes.\\ 

The results show that hydrogen uptake is enhanced through two mechanisms: (i) higher hydrogen reaction rates due to lower applied potentials, and (ii) a smaller pH resulting from corrosion, as observed at high applied potentials. Accordingly, there is an intermediate regime where the hydrogen uptake is reduced. This is shown in Fig. \ref{fig:Em_LatticeH}, where crack tip predictions of pH (blue crosses, left $y$-axis) and lattice hydrogen concentration $C_L$ (orange circles, right $y$-axis) are shown as a function of the applied potential. A point of minimum hydrogen uptake is observed at $E_m=-0.2 \mathrm{V}_{SHE}$ ($-0.441 \mathrm{V}_{SCE}$). Such behaviour is also observed experimentally, with hydrogen embrittlement susceptibility diminishing with increasing applied potential up to a certain point, after which susceptibility increases with $E_m$ \cite{AM2016}. Further insight into the dependence on the applied potential and the competition between the different reaction kinetics can be gained through Fig. \ref{fig:Em_ReactionRate}, where individual reaction rates are reported as a function of $E_m$. As predicted from Eq. \eqref{eq:scaling_15}, reaction $\nu_{Va}$ is dominant for low pH values, while $\nu_{Vb}$ becomes the dominant reaction for pH values above 7. The application of a negative metal potential increases the overpotential to accelerate reaction $\nu_{Vb}$, from being almost negligible to becoming the sole source of adsorbed hydrogen. In contrast, positive electric potentials slow down all hydrogen reactions by altering the overpotential, while accelerating reaction $\nu_{Va}$ by strongly decreasing the pH. This causes a strong rise in lattice hydrogen going from neutral to positive potentials. However, this rise flattens for higher potentials due to the increased availability of $\mathrm{H}^+$ being negated by the reduction of the reaction rate through the overpotential. 

\subsection{Fluid velocity}
\label{sec:fluid_V}

\begin{figure}
    \centering
    \begin{subfigure}{12cm}
         \centering
         \includegraphics[width=12cm]{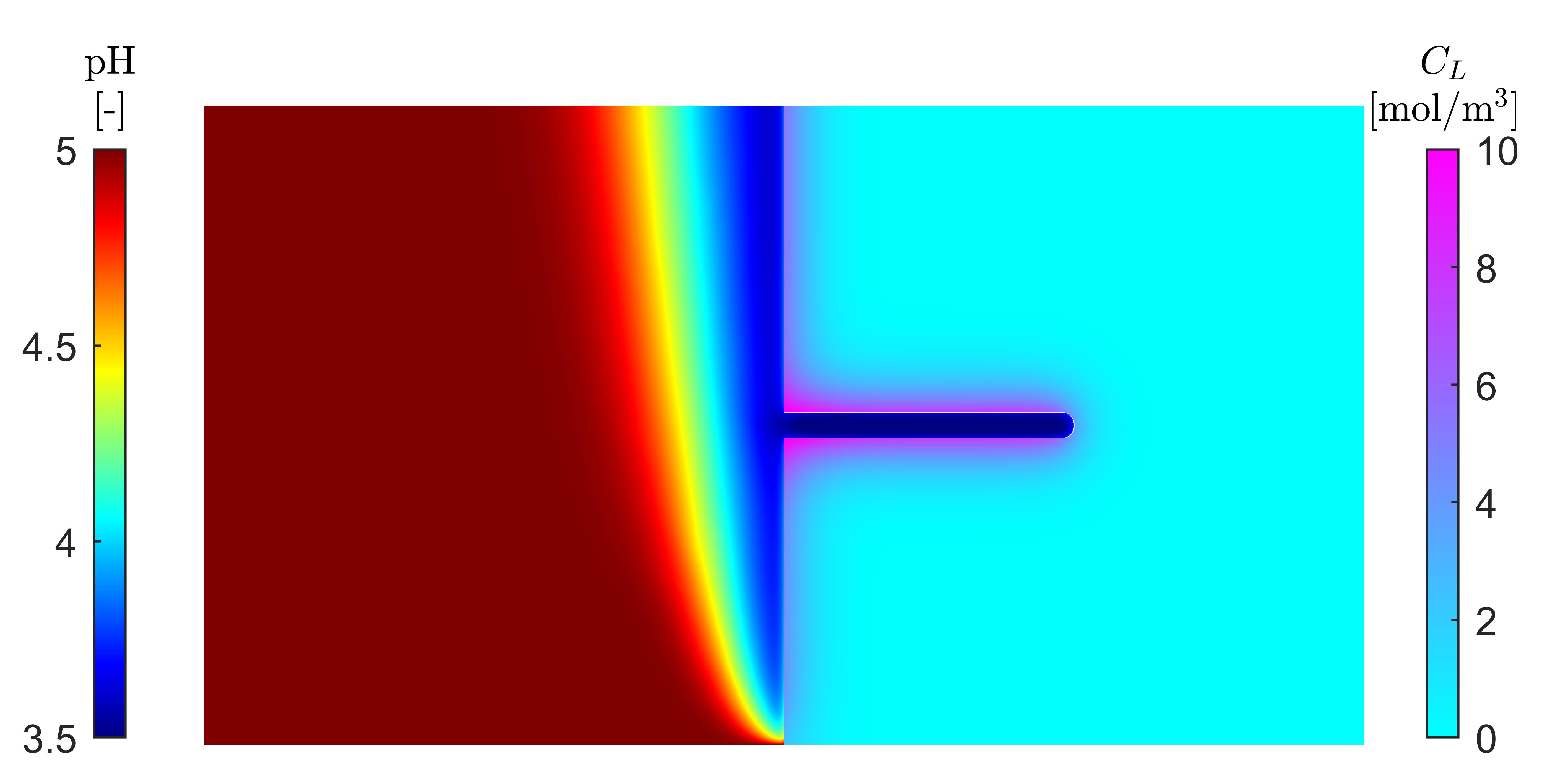}
         \caption{$V_{max}=1\;\mathrm{mm}/\mathrm{s}$}
         \label{fig:Vel1}
    \end{subfigure}
    \begin{subfigure}{12cm}
         \centering
         \includegraphics[width=12cm]{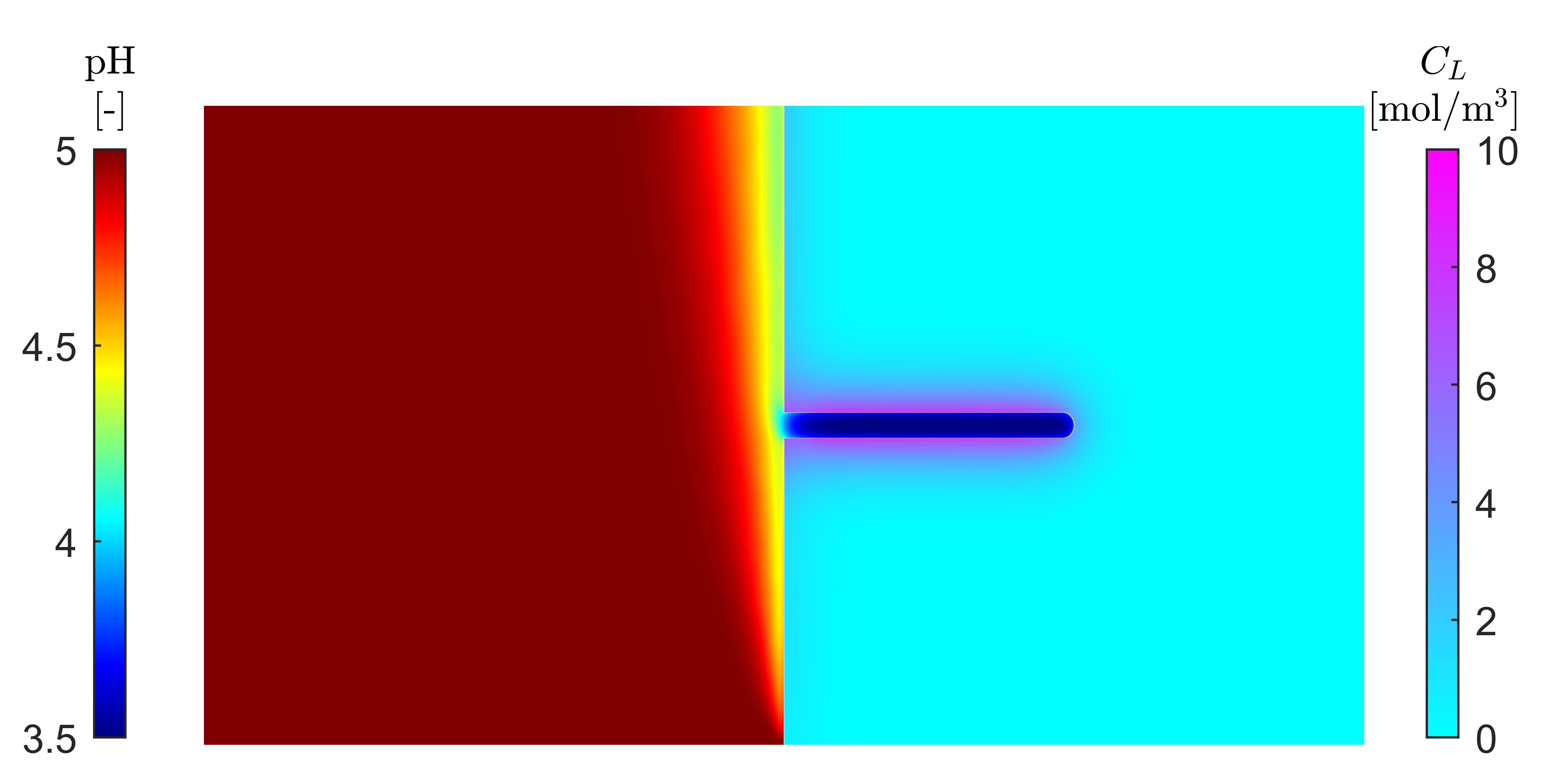}
         \caption{$V_{max}=20\;\mathrm{mm}/\mathrm{s}$}
         \label{fig:Vel20}
     \end{subfigure}
    \caption{Metal-electrolyte interactions: Effect of the electrolyte velocity on the pH and lattice hydrogen concentration at $t=5\;\mathrm{min}$. The electrolyte velocity varies linearly from $V_{max}$ at the left edge of the electrolyte, to 0 at the electrolyte-metal interface (see Fig. \ref{fig:geo}). Here, results are provided for the choices of (a) $V_{max}=1\;\mathrm{mm}/\mathrm{s}$ and (b) $V_{max}=20\;\mathrm{mm}/\mathrm{s}$.}
    \label{fig:Vel}
\end{figure}

\begin{figure}
    \centering
    \includegraphics{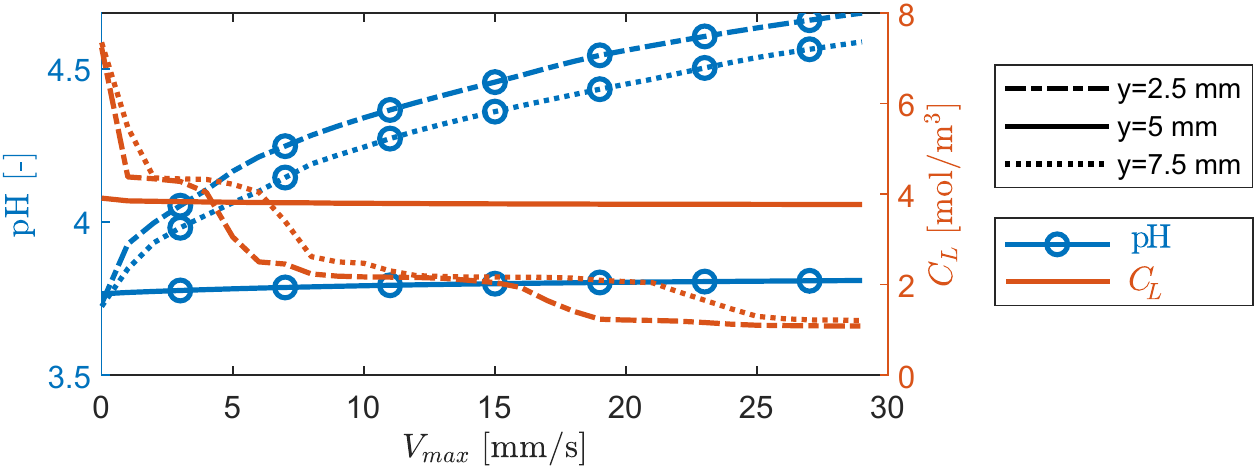}
    \caption{Metal-electrolyte interactions: Effect of the electrolyte velocity on the pH (blue, left $y$-axis) and lattice hydrogen concentration (orange, right $y$-axis) at $t=5\;\mathrm{min}$ and various locations along the metal-electrolyte interface ($y$, with $y=5$ mm being the centre height).}
    \label{fig:Vel_LatticeH}
\end{figure}

While it is accurate to assume negligible fluid flow within the occluded geometry of the crack, this is less realistic for the electrolyte located in the bulk of the domain. To investigate the interplay between the bulk electrolyte velocity and hydrogen ingress, we prescribe a fluid velocity that varies linearly from a magnitude $V_{max}$ at the edge of the electrolyte to zero at the electrolyte-metal interface (see Fig. \ref{fig:geo}). Specifically, we vary the maximum velocity from $V_{max}=0\;\mathrm{mm}/\mathrm{s}$ (corresponding to the results from the previous section) up to $V_{max}=29\;\mathrm{mm}/\mathrm{s}$, using  $1\;\mathrm{mm}/\mathrm{s}$ intervals. \\

The distributions of electrolyte pH and metal lattice hydrogen concentration are shown in Fig. \ref{fig:Vel} for the representative cases of $V_{max}=1\;\mathrm{mm}/\mathrm{s}$ and $V_{max}=20\;\mathrm{mm}/\mathrm{s}$. The locations at which the $\mathrm{pH}$ and lattice concentration are shown are the crack tip ($y=5\;\mathrm{mm}$), and at the exterior surface $2.5\;\mathrm{mm}$ above and below the crack mouth ($y=2.5\;\mathrm{mm}$ and $y=7.5\;\mathrm{mm}$). The results show that, while a moving electrolyte has a significant effect on the concentrations within the bulk and near the exterior of the metal, the pH inside the defect is rather insensitive. This independence of the local pH on the bulk electrolyte velocity is due to the fact that the electrolyte near the crack tip is in a state of local equilibrium, as it is located too far from the outer domain for any meaningful quantity of ions to diffuse to the crack tip. As a result, the hydrogen uptake near the crack tip within the metal also shows only a small sensitivity to the electrolyte velocity. However, the effect of the electrolyte velocity on the hydrogen uptake near the exterior boundary is more noticeable. This is shown in Fig. \ref{fig:Vel_LatticeH}, where the pH (left, blue colour) and the lattice hydrogen concentration (right, orange) are plotted as a function of $V_{max}$ at different positions along the interface height ($y$). Outside of the crack mouth, the pH is sensitive to the fluid velocity, increasing with $V_{max}$ up to the point of becoming closer to its initial value (pH=5). While corrosion decreases the pH locally, this effect is limited for higher velocities where the $\mathrm{Fe}^{2+}$ and $\mathrm{FeOH}^+$ ions are removed due to advection before they can react to create $\mathrm{H}^+$ ions. Furthermore, the imposed velocity removes the $\mathrm{H}^+$ ions that are added as a result of the corrosion reaction. These ions are advected upwards along the metal-electrolyte interface, causing the ion concentrations nearer to the bottom of the electrolyte domain to be close to the boundary conditions. Higher up, the combination of advected ions and newly created ions due to reactions causes the pH to deviate more from these boundary conditions. As a result, the effect of including the fluid flow on the pH is strongest near the bottom of the domain (compare with the $V_{max}=0$ result in Fig. \ref{fig:Em}), whereas its effect is lessened further upward. The fluid velocity also changes the electric potential near the metal-electrolyte interface, causing higher potentials closer to the bottom of the domain, while lower potentials are observed nearer the top. Hence, the electric potential can locally counteract the effect of pH on hydrogen ingress. Thus, while a higher fluid velocity leads to a raise in bulk pH near the interface (resulting in less hydrogen absorption in the exterior boundaries), the increase in $\varphi$ with fluid velocity in the bottom region of the domain results in a reduction in the hydrogen uptake close to the inflow boundary condition. This indicates that if the fluid velocity is sufficiently high compared to the domain size, simulating the exterior electrolyte becomes less relevant, and simply imposing the initial concentrations might be appropriate. However, this is not valid for occluded regions, where the pH is significantly different from the initial concentration and insensitive to the imposed velocity. 

\subsection{Defect geometry}
\label{sec:res_geo}
\begin{figure}
    \centering
    \begin{subfigure}{8cm}
         \centering
         \includegraphics{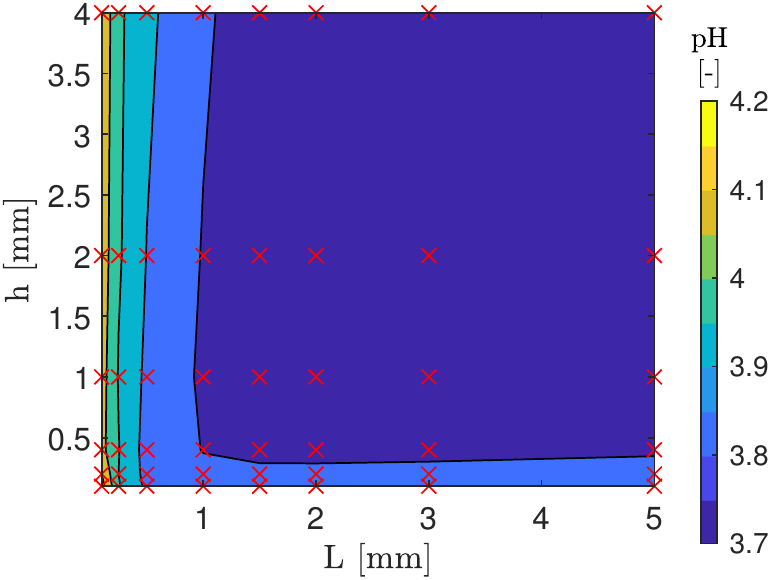}
         \caption{}
         \label{fig:hL_surfs_pH}
     \end{subfigure}
    \begin{subfigure}{8cm}
         \centering
         \includegraphics{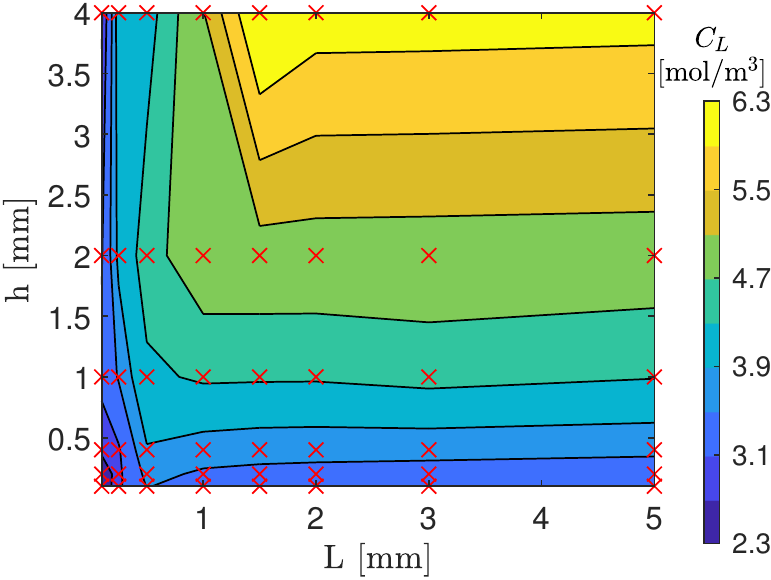}
         \caption{}
         \label{fig:hL_surfs_CL}
    \end{subfigure}
    \begin{subfigure}{8cm}
         \centering
         \includegraphics{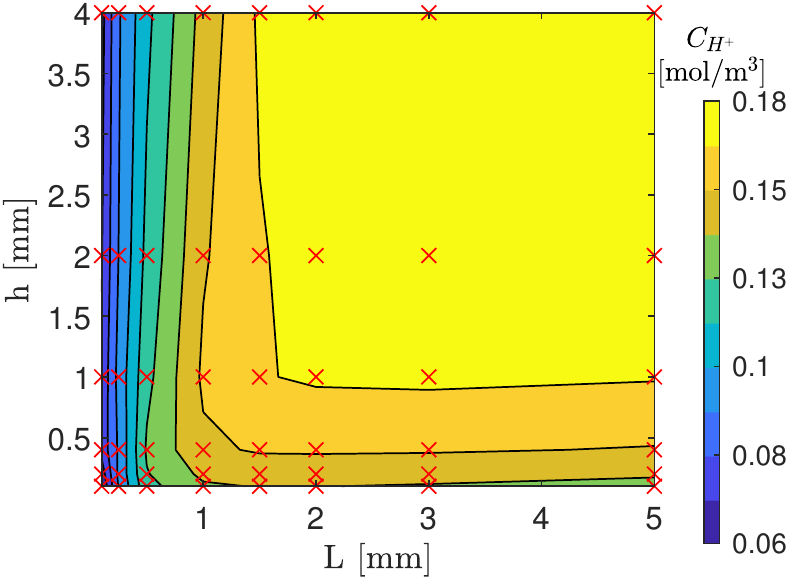}
         \caption{}
         \label{fig:hL_surfs_H}
    \end{subfigure}
    \begin{subfigure}{8cm}
         \centering
         \includegraphics{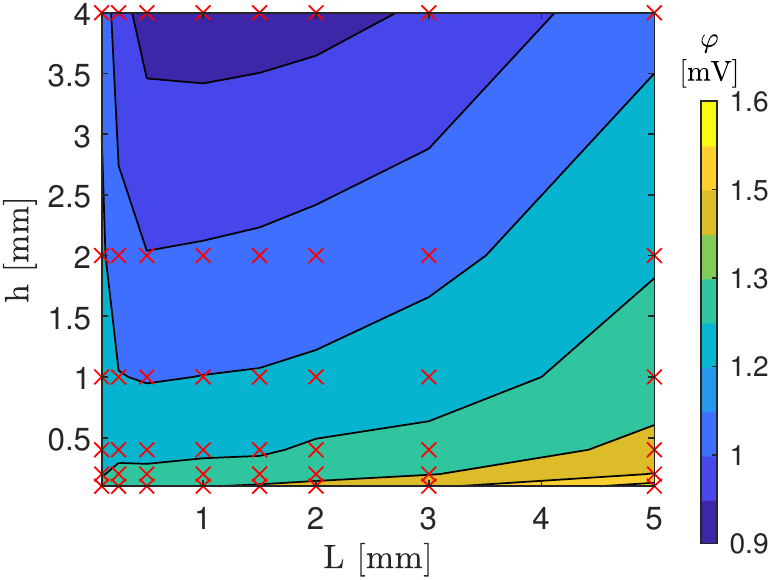}
         \caption{}
         \label{fig:hL_surfs_E}
     \end{subfigure}
    \caption{Metal-electrolyte interactions: effect of the defect geometry on crack tip (a) pH, (b) absorbed lattice hydrogen concentration $C_L$, (c) $\mathrm{H}^+$ concentration, and (d) electrolyte electric potential $\varphi$. The maps are built for a range of defect lengths ($L=0.1$ mm to $L=5$ mm) and heights ($h=0.1$ mm to $h=4$ mm). Results shown at $t=5\;\mathrm{mins}$, with the red crosses indicating simulation data points and the contours being built by interpolating linearly between these points. }
    \label{fig:hL_surfs}
\end{figure}

\begin{figure}
    \centering
    \begin{subfigure}{12cm}
         \centering
         \includegraphics[width=12cm]{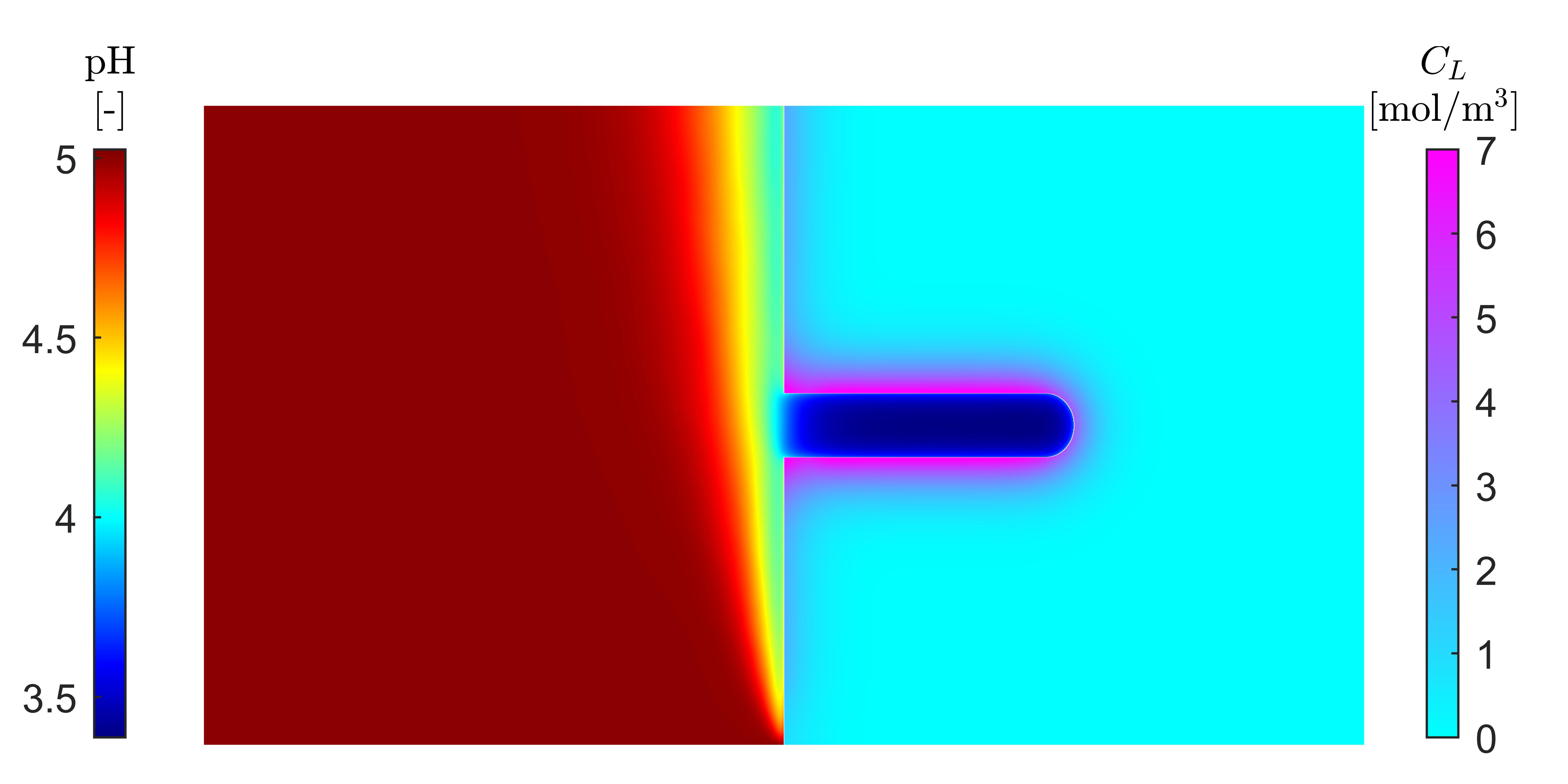}
         \caption{$h=1\;\mathrm{mm}$}
         \label{fig:LH_opening1}
    \end{subfigure}
    \begin{subfigure}{12cm}
         \centering
         \includegraphics[width=12cm]{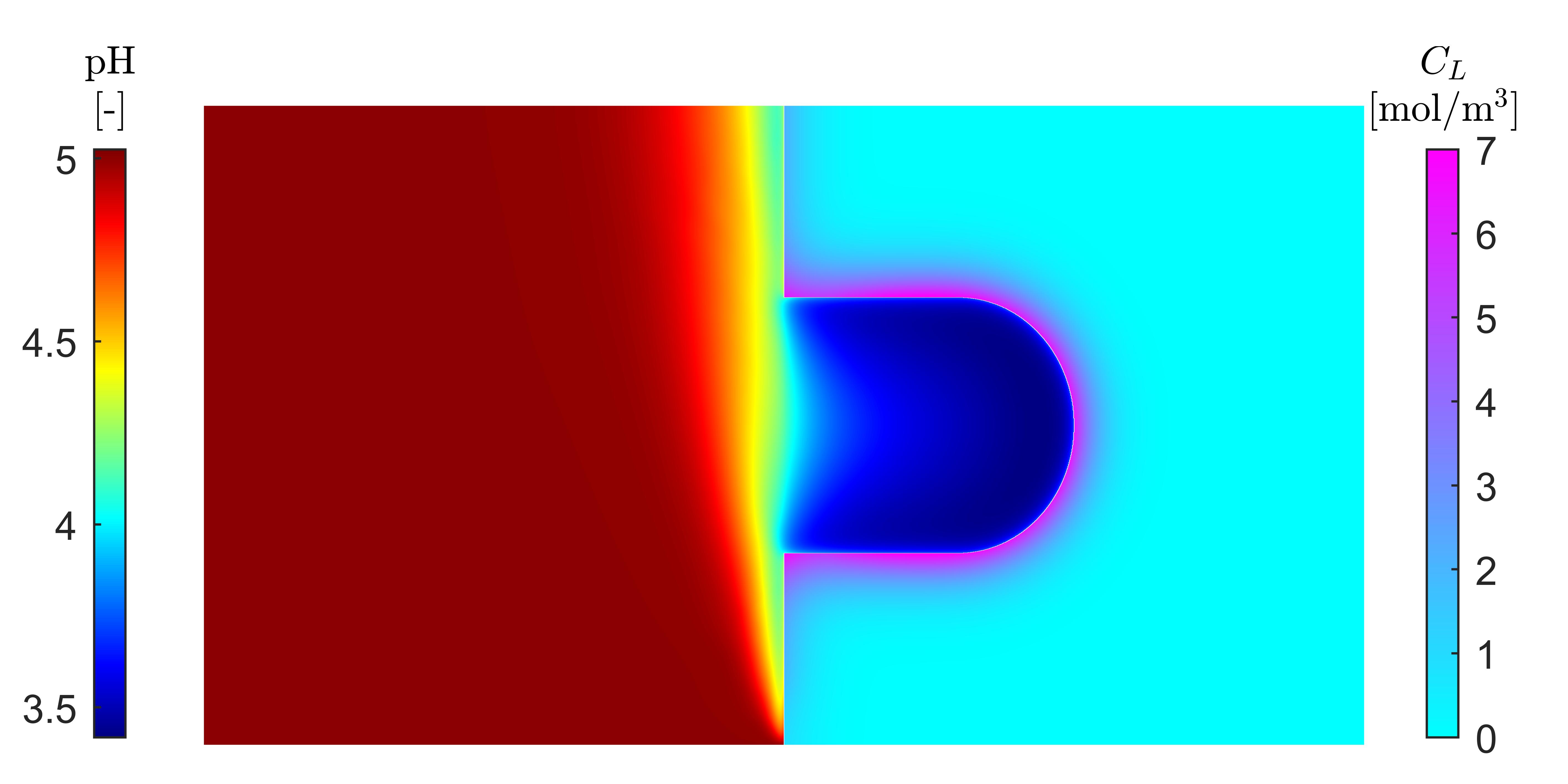}
         \caption{$h=4\;\mathrm{mm}$}
         \label{fig:LH_opening4}
     \end{subfigure}
    \caption{Metal-electrolyte interactions: influence of the defect dimensions. Electrolyte pH and absorbed lattice hydrogen concentration in the metal for two representative case studies: (a) a defect of height $h=1\;\mathrm{mm}$, and (b) a defect of height $h=5\;\mathrm{mm}$. In both cases, the defect length equals $L=5\;\mathrm{mm}$, and the results correspond with a time of $t=5\;\mathrm{min}$.}
    \label{fig:LH_opening}
\end{figure}

As seen in the previous section, the pH and the hydrogen uptake near the crack tip are not influenced by the exterior electrolyte's behaviour and pH when the crack is sufficiently long. This was also seen in the scaling analysis of Eq. \eqref{eq:scaling_LH}, indicating that for all but the shortest crack lengths the acidic hydrogen evolution reaction rate will be limited by the hydrogen diffusion into the crack and by the hydrogen produced by local corrosion reactions. In this section, we will investigate the effect of the defect geometry, changing the defect length over a range going from $L=0.1$ mm to $L=5$ mm and the defect height from $h=0.1$ mm to $h=4$ mm. The radius at the defect tip is taken as the smaller of the two dimensions. This span of $L$ and $h$ values aims at covering a wide range of occluded geometries, from pit-like circular defects that can arise from localised corrosion to long sharp cracks. For the fluid velocity, we will use $V_{max}=10\;\mathrm{mm}/\mathrm{s}$, enforcing a near to constant pH at the defect mouth. The results obtained are shown in Fig. \ref{fig:hL_surfs}, where maps are constructed that relate the defect geometry to the crack tip estimates of pH (Fig. \ref{fig:hL_surfs}a), absorbed lattice hydrogen (Fig. \ref{fig:hL_surfs}b), $\mathrm{H}^+$ concentration (Fig. \ref{fig:hL_surfs}c), and electrolyte potential (Fig. \ref{fig:hL_surfs}d).\\

Consider first the pH and $\mathrm{H}^+$ concentration results, Figs. \ref{fig:hL_surfs}a and \ref{fig:hL_surfs}c. For shorter defects ($L<2\;\mathrm{mm}$), the pH shows a strong sensitivity to the defect length but only a weak sensitivity to the defect opening. The $\mathrm{H}^+$ generated through the reacting $\mathrm{Fe}^{2+}$ diffuses more readily away from the defect tip when the defect length is small. Hence, wider pits allow for more diffusion compared to their surface area, resulting in a pH closer to the exterior conditions. For defects longer than $2\;\mathrm{mm}$, diffusion becomes negligible and the pH is solely governed by the local equilibrium of the hydrogen and corrosion reactions. This strong dependency on the defect length, and weak dependency on its height, corresponds roughly with the predictions from Eq. \eqref{eq:scaling_LH}, as discussed in Section \ref{sec:dimensionalAnalysis}. However, it should be noted that what determines if the results are dominated by the local reactions or by diffusion for a near-neutral metal potential is the diffusion of $\mathrm{H}^+$ ions away from the defect, and not towards it. Consider now the sensitivity of crack tip electrolyte potential to the defect dimensions, Fig. \ref{fig:hL_surfs}d. As it can be observed, the crack tip value of $\varphi$ generally increases with the defect length. Longer and thinner defects hinder electrolyte ionic transport and lead to noticeable increases in the crack tip electrolyte potential. Finally, the uptake of hydrogen is shown in Fig. \ref{fig:hL_surfs_CL}, in terms of the absorbed lattice hydrogen concentration at the crack tip. The results reveal a significant sensitivity to the defect geometry. In particular, the lattice hydrogen content increases with the defect height. This is also observed for large defects, where changes in the pH and electrolyte potential are low. This behaviour is intrinsic to the crack tip geometry as the hydrogen diffuses away from it into the metal. For narrow cracks with small crack tip radii, the radial diffusion of $C_L$ implies that the diffusion will spread the available lattice hydrogen over a relatively large area. In contrast, defects with a large opening height and a large tip radius, such as pits, will cause almost one-dimensional diffusion away from their tip. This effect is shown in Fig. \ref{fig:LH_opening}, where contours of electrolyte pH and metal lattice hydrogen content are shown for two selected values of the defect height. The hydrogen concentrations at the top and bottom faces of the crack are almost identical, as could be expected given the similar pH and low electric potential, while the hydrogen around the crack tip spreads over a larger region for the $1\;\mathrm{mm}$ height (relative to the $4\;\mathrm{mm}$ case). However, these bulk diffusion effects can be naturally captured by existing metal deformation-diffusion models, without explicitly simulating the electrolyte. \\

In terms of electrolyte-geometry interactions, our results show that when the defect is sufficiently long ($L>2\;\mathrm{mm}$), the defect geometry is less relevant, as long as the influence of the electric potential is low. For these cases, having an exact description of the geometry is not needed to estimate the local pH, and one can consider local equilibrium to describe the environmental conditions. Similarly, if the local environmental conditions are known for a specific geometry, it can be reasonably expected that they would be applicable to other geometries within the ``large crack'' regime ($L>2\;\mathrm{mm}$, for the material and conditions considered here). In contrast, for shorter defects the geometry has a significant effect on electrolyte behaviour, as the diffusion of ions into and out of the defect becomes important. In those circumstances, small deviations in crack length can cause significant changes in environmental conditions and hydrogen uptake.

\FloatBarrier
\section{Assessment of modelling strategies}
\label{sec:bcs}

\begin{figure}
    \centering
    \includegraphics{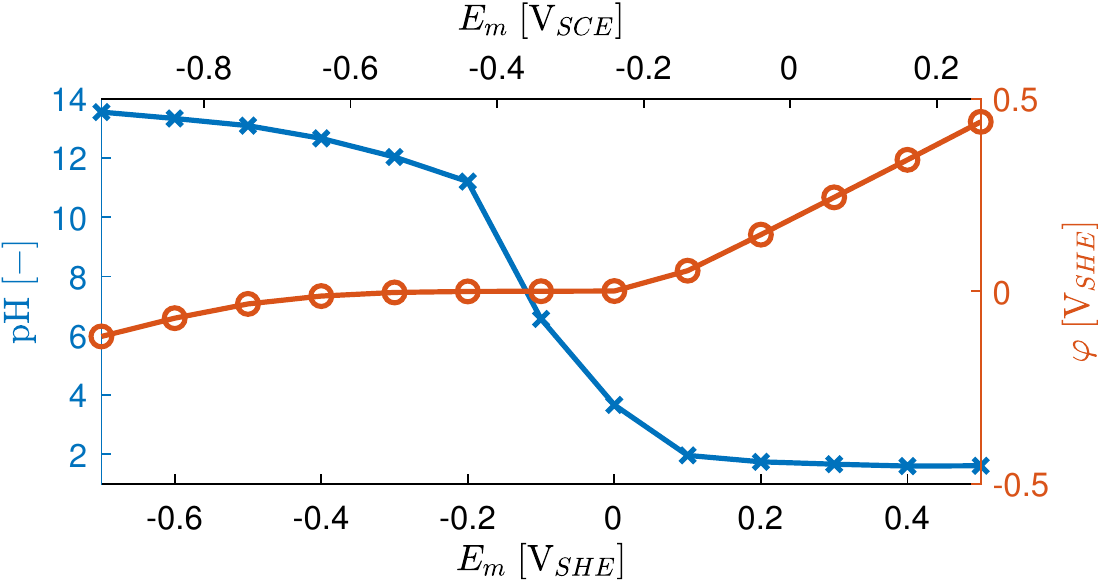}
    \caption{Mapping the crack tip pH and electrolyte potential as a function of the applied potential $E_m$. The results correspond for the case of a sufficiently large crack, after a time of $t=10\;\mathrm{min}$, neglecting the fluid velocity and prescribing a pH of 5 and a potential of $\varphi=0\;\mathrm{V}_{SHE}$ at the external electrolyte edge (located at $10\;\mathrm{mm}$ from the metal surface). The pH results use blue crosses and refer to the left $y$-axis, while the electrolyte potential results use orange circles and refer to the right $y$-axis.}
    \label{fig:EnvironmentalMap}
\end{figure}

The results shown so far reveal that the pH and the electrolyte potential near the metal surface deviate significantly from the initial pH and the applied potential. However, the results presented in Section \ref{sec:res_geo} also show that the electric potential is only weakly dependent on the crack geometry, and that an almost constant pH is obtained inside of the defect, as long as the defect is sufficiently long. Hence, our model can be used to determine the local pH and electrolyte potential associated with a given applied potential, and these would be relevant for all sufficiently long cracks. These relations, shown in Fig. \ref{fig:EnvironmentalMap}, can be used as input to simplified hydrogen uptake models that can provide a relatively accurate quantification of hydrogen ingress without the need to resolve the complete electro-chemo-mechanical problem. Thus, we proceed to derive simplified relationships and assess their accuracy, as well as to inspect the predictions from simplistic yet widely-used models. As a word of caution, one should note that the environmental maps such as the one provided in Fig. \ref{fig:EnvironmentalMap}, which can serve as key input to simplified models, are solely an estimate and do not capture the initial period during which the pH near the fracture tip slowly changes towards a more stable solution. They are also not representative of the steady-state behaviour obtained when the metal lattice is fully saturated with hydrogen. However, in this intermediate period, they can give a sensible estimate for the environmental conditions based on the applied metal potential. Throughout this section, the results refer to an Fe-based material, as characterised by the parameters shown in Tables \ref{table:params} and \ref{tab:reactionsused}.\\

Building upon local environmental maps such as the one provided in Fig. \ref{fig:EnvironmentalMap}, let us proceed to derive simplified estimates of the hydrogen influx. First, assuming the number of surface sites to be small, and hence the surface to be in a state of local equilibrium, Eq. \eqref{eq:massbalanceinterface} can be simplified to:
\begin{equation}
    J = \nu_A-\nu_A'= (\nu_{Va}-\nu_{Va}') - (\nu_{Ha}-\nu_{Ha}') -2 (\nu_T-\nu_T')  + (\nu_{Vb}-\nu_{Vb}') - (\nu_{Hb}-\nu_{Hb}')
    \label{eq:massbalanceinterface_steady}
\end{equation}
Next, we neglect all backwards reaction rates except for $\nu_A'$. This is  sensible as long as these backwards reaction rates are sufficiently small compared to their forward rates. The implication is that the hydrogen is solely adsorbed through the Volmer or backwards absorption reactions, and that is solely removed from the metal surface through the Tafel, Heyrovsky, and forwards absorption reactions. As a result, the interfacial mass balance dictates the hydrogen influx into the metal as:
\begin{equation}
\begin{split}
    J = \nu_A-\nu_A' = &k_{Va} C_{\mathrm{H}^+} (1-\theta_{ads}) \exp{\left(-\alpha_{Va} \frac{\eta F}{RT}\right)} - k_{Ha} C_{\mathrm{H}^+} \theta_{ads} \exp{\left(-\alpha_{Ha} \frac{\eta F}{RT}\right)} \\ &-2k_T\theta_{ads}^2+k_{Vb} (1-\theta_{ads})\exp{\left(-\alpha_{Vb} \frac{\eta F}{RT}\right)} - k_{Hb} \theta_{ads} \exp{\left(-\alpha_{Hb} \frac{\eta F}{RT}\right)}
    \end{split}\label{eq:49}
\end{equation}

\begin{figure}
    \centering
    \includegraphics{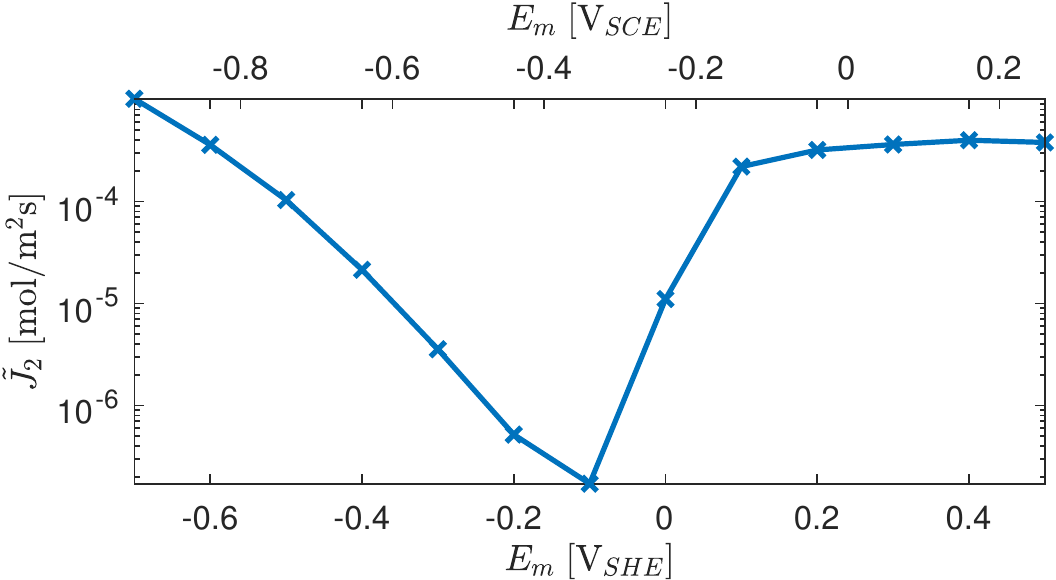}
    \caption{Hydrogen entry rate estimated from the simplified model shown in 
    Eq. (\ref{eq:env_simple}) and the environmental conditions given in Figure \ref{fig:EnvironmentalMap}. The results correspond for the case of a sufficiently large crack, after a time of $t=10\;\mathrm{min}$, neglecting the fluid velocity and prescribing a pH of 5 and a potential of $\varphi=0\;\mathrm{V}_{SHE}$ at the external electrolyte edge (located at $10\;\mathrm{mm}$ from the metal surface). Other parameters used $k_{Va}=10^{-4}\;\mathrm{m}/\mathrm{s}$, $k_{Vb}=10^{-8}\;\mathrm{m}/\mathrm{s}$ and $\alpha = 0.5$.}
    \label{fig:EnvironmentalMap_Eval}
\end{figure}

\begin{figure}[t]
    \centering
    \begin{subfigure}{6cm}
         \centering
         \includegraphics{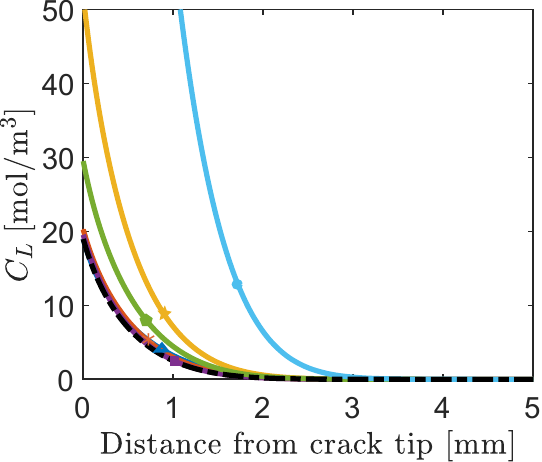}
         \caption{}
         \label{fig:BC-05a}
    \end{subfigure}
    \begin{subfigure}{10cm}
         \centering
         \includegraphics{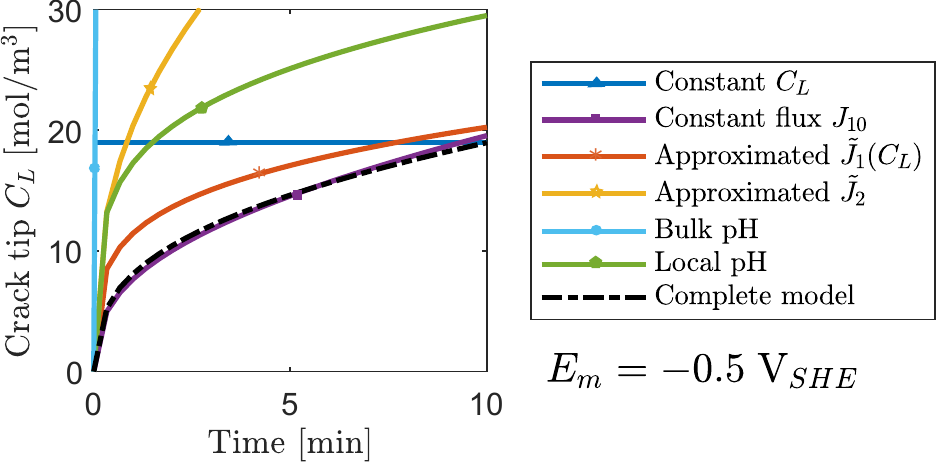}
         \caption{\hspace{3cm}\;}
         \label{fig:BC-05b}
     \end{subfigure}
\\
    \centering
    \begin{subfigure}{6cm}
         \centering
         \includegraphics{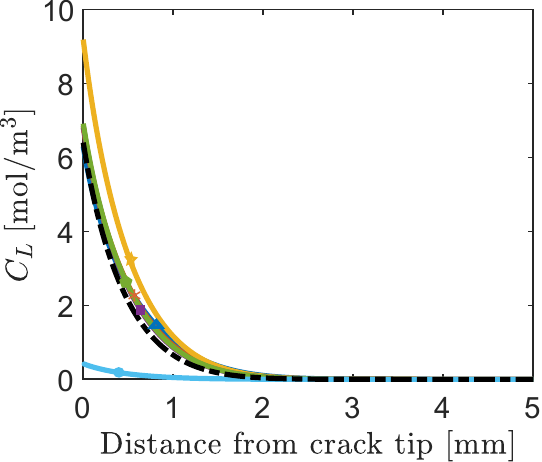}
         \caption{}
         \label{fig:BC0a}
    \end{subfigure}
    \begin{subfigure}{10cm}
         \centering
         \includegraphics{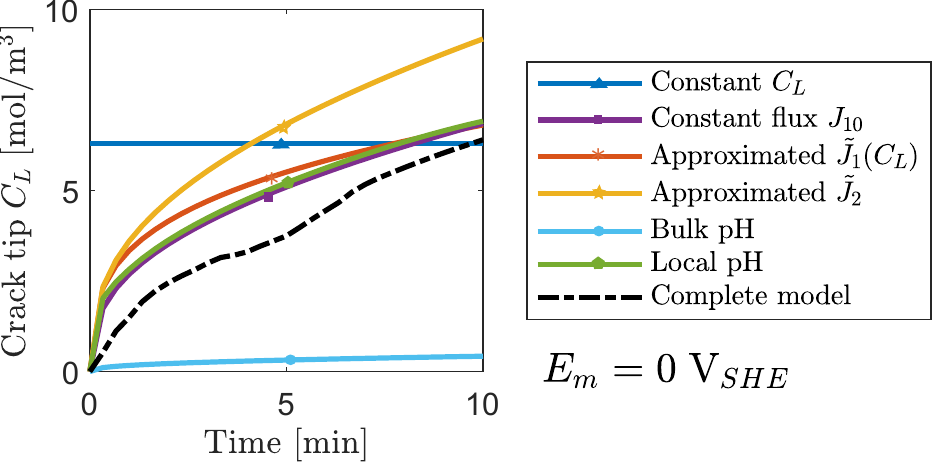}
         \caption{\hspace{3cm}\;}
         \label{fig:BC0b}
     \end{subfigure}
\\
    \begin{subfigure}{6cm}
         \centering
         \includegraphics{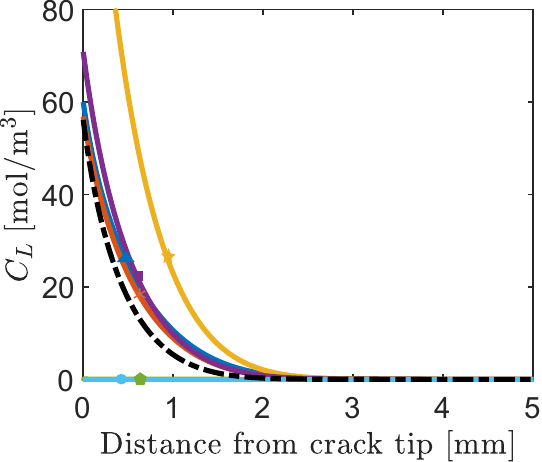}
         \caption{}
         \label{fig:BC05a}
    \end{subfigure}
    \begin{subfigure}{10cm}
         \centering
         \includegraphics{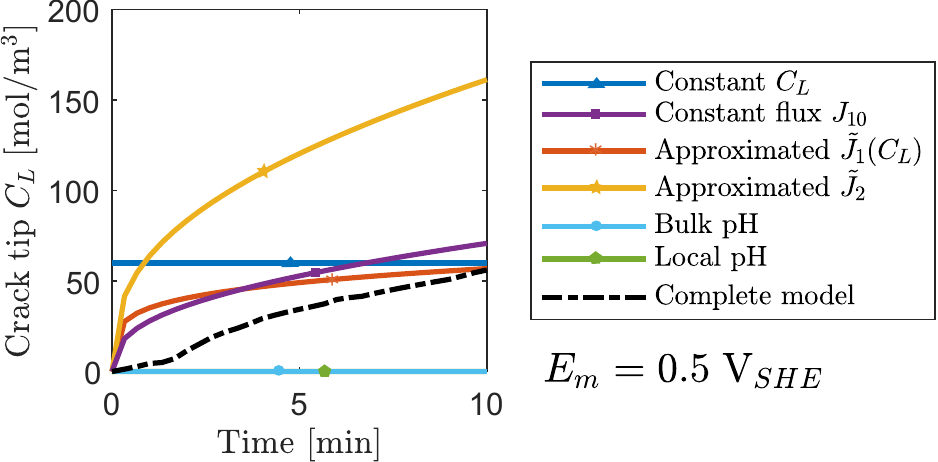}
         \caption{\hspace{3cm}\;}
         \label{fig:BC05b}
     \end{subfigure}
    \caption{Assessment of modelling strategies. Lattice hydrogen distribution ahead of the crack after $t=10$ min. (left) and evolution of the crack tip lattice hydrogen concentration (right). Seven modelling strategies are considered: (i) prescribing a constant hydrogen concentration, (ii) prescribing a constant hydrogen flux, (iii) our approximated flux model $\tilde{J}_1$, (iv) our further simplified flux model $\tilde{J}_2$, two cases where the electrochemistry is not solved for upon the assumption that the pH is known, one using the bulk pH (v) and another one using the local one (vi), and (vii) our complete electro-chemo-mechanical model (the reference result). Results are shown for three values of the applied potential: $E_m = -0.5\;\mathrm{V}_{SHE}$ (a \& b), $E_m = 0\;\mathrm{V}_{SHE}$ (c \& d) and $E_m = 0.5\;\mathrm{V}_{SHE}$ (e \& f). Based on the assumptions discussed and Fig. \ref{fig:EnvironmentalMap_Eval}, the values used are: $C_L=19\;\mathrm{mol}/\mathrm{m}^3$, $\tilde{J}_2=1.1\cdot10^{-4}\;\mathrm{mol}/(\mathrm{m^2}\;\mathrm{s})$, $J_{10}=4.2\cdot10^{-5}\;\mathrm{mol}/(\mathrm{m^2}\;\mathrm{s})$, and local pH$=13$ ($E_m = -0.5\;\mathrm{V}_{SHE}$); $C_L=6.3\;\mathrm{mol}/\mathrm{m}^3$, $\tilde{J}_2=2.0\cdot10^{-5}\;\mathrm{mol}/(\mathrm{m^2}\;\mathrm{s})$, $J_{10}=1.5\cdot10^{-5}\;\mathrm{mol}/(\mathrm{m^2}\;\mathrm{s})$, and local pH$=3.7$ ($E_m = 0\;\mathrm{V}_{SHE}$); $C_L=60\;\mathrm{mol}/\mathrm{m}^3$, $\tilde{J}_2=3.4\cdot10^{-4}\;\mathrm{mol}/(\mathrm{m^2}\;\mathrm{s})$, $J_{10}=1.5\cdot10^{-4}\;\mathrm{mol}/(\mathrm{m^2}\;\mathrm{s})$, and local pH$=1.6$ and ($E_m = 0.5\;\mathrm{V}_{SHE}$).}
    \label{fig:BC}
\end{figure}

A further simplification is made, assuming the absorption reaction to occur much faster compared to all other reactions, resulting in the surface and lattice hydrogen concentrations to be in local equilibrium, and related to each other through:
\begin{equation}
    \theta_{ads} = \frac{C_L}{\frac{k_A}{k_A'}(N_L-C_L)+C_L}
\end{equation}
As a result, the environmental conditions ($C_{\mathrm{H}^+}$ and $\varphi$) and current hydrogen concentration within the metal can be used to impose an approximate hydrogen influx as:
\begin{align}
\begin{split}
       &\tilde{J}_1=  \left(1-\frac{C_L}{\frac{k_A}{k_A'}(N_L-C_L)+C_L} \right) \left(k_{Va} C_{\mathrm{H}^+}  \exp{\left(-\alpha_{Va} \frac{(E_m-E_{eq}-\varphi) F}{RT}\right)} + k_{Vb} \exp{\left(-\alpha_{Vb} \frac{(E_m-E_{eq}-\varphi) F}{RT}\right)}\right) \\ &- \frac{C_L}{\frac{k_A}{k_A'}(N_L-C_L)+C_L} \Bigg( k_{Ha} C_{\mathrm{H}^+} \exp{\left(-\alpha_{Ha} \frac{(E_m-E_{eq}-\varphi) F}{RT}\right)}  \\ & + k_{Hb} \exp{\left(-\alpha_{Hb} \frac{(E_m-E_{eq}-\varphi) F}{RT}\right)}  +2k_T \frac{C_L}{\frac{k_A}{k_A'}(N_L-C_L)+C_L} \Bigg)  
\end{split} 
\label{eq:mapping}
\end{align}
Under the assumptions outlined above, this equation can be used to impose a hydrogen influx \textit{via} a non-linear Neumann-type boundary condition. At this point, it is of interest to compare this expression to the state-of-the-art generalised flux boundary condition for hydrogen ingress; namely, that of Turnbull and co-workers \cite{Turnbull1996,Turnbull2015,CS2020}. The main differences are the following; in our simplified model $\tilde{J}_1$, (i) non-acidic reactions are not neglected, and (ii) the effects of $\mathrm{H}^+$ and $\mathrm{OH}^-$ concentrations are not encapsulated in the reaction constants, ensuring the applicability of a single set of constants over a wide range of environmental conditions. Thus, the approximation from Eq. \eqref{eq:mapping} is valid over a larger range of external environments and is able to accommodate environment-independent reaction constants.\\

One further simplification can be made for relatively short time scales; assuming $C_L<<N_Lk_A/k_A'$ (for our reaction constants, valid up to $C_L\approx 10^2\;\mathrm{mol}/\mathrm{m}^3\approx 13\;\mathrm{wppm}$) and assuming that the electrolyte potential is given relative to the hydrogen reactions $E_{eq}=0$. This reduces Eq. \eqref{eq:mapping} to:
\begin{equation}
    \tilde{J}_{2} = \left(k_{Va} 10^{-pH+3}+k_{Vb}\right) \exp{\left(-\alpha \frac{(E_m-\varphi) F}{RT}\right)} 
    \label{eq:env_simple}
\end{equation}

Eq. (\ref{eq:env_simple}) provides a straightforward relationship between the environment (pH, $\varphi$, $E_m$) and the hydrogen influx. For the case of a long crack, and the reaction constants and conditions considered here, the resulting hydrogen influx is given in Fig. \ref{fig:EnvironmentalMap_Eval} as a function of the applied potential. Thus, the information given in Fig. \ref{fig:EnvironmentalMap_Eval} can be used as input to standard chemo-mechanical models that do not resolve the electrochemistry, upon the assumptions discussed above.\\ 

In the remainder of this section, we will compare the results obtained with the complete electro-chemo-mechanics model to those obtained from various simplified models, so as to assess their accuracy. In particular, in addition to the reference case (the `complete' model), we consider: (i) the most commonly used approach of prescribing a constant hydrogen concentration, (ii) prescribing a constant hydrogen flux, (iii) our approximated flux model $\tilde{J}_1$, (iv) our further simplified flux model $\tilde{J}_2$, and two cases where the electrochemistry is not solved for upon the assumption that the pH is known, one using the bulk pH (v) and another one using the local one (vi). It should be noted that current models that prescribe a constant hydrogen concentration are unable to relate the environment to the magnitude of $C_L$ prescribed in the crack faces. Here, we assume that one has access to a map such as the one provided in Fig. \ref{fig:hL_surfs}b, and take that input from a complete electro-chemo-mechanical model as the boundary value of $C_L$. So, the value of $C_L$ is chosen to match the reference result at a particular time (here, $t=10$ min.) and then its evolution is assessed. Also, the constant hydrogen flux, case (ii), is chosen assuming access to the outcome of a complete electro-chemo-mechanical simulation. Specifically, the flux prescribed corresponds to the flux obtained with the complete model at a certain time  (again, $t=10$ min., $J_{10}$). The magnitude of $\tilde{J}_1$ is estimated from (\ref{eq:mapping}), taking the crack tip pH and electrolyte potential from the map provided in Fig. \ref{fig:EnvironmentalMap_Eval} and from the current value of $C_L$ (a primary variable of the model). And the pH-based approaches, (v) and (iv), solve both the deformation-diffusion problem in the metal and the surface kinetics ($\theta_{ads}$), but do not resolve the electrolyte electrochemistry problem - assuming $\varphi=0$ and, for the case of the local pH, estimating this through the map provided in Fig. \ref{fig:EnvironmentalMap_Eval}. Hence, the boundary conditions of strategies (iii), (v), (vi) and the complete electro-chemo-mechanical model are time- and solution-dependent; unlike modelling strategies (i), (ii) and (iv), which take a constant $C_L$ or $J$ value. Calculations are conducted for three selected values of the applied potential: $E_m=-0.5$, $0$, and $0.5\;\mathrm{V}_{SHE}$. For $E_m=0.5\;\mathrm{V}_{SHE}$ it follows from Fig. \ref{fig:EnvironmentalMap} that a good approximation for the local pH and electrolyte potential are 1.6 and $0.4\;\mathrm{V}_{SHE}$, respectively. Based on Eq. \eqref{eq:env_simple}, this results in $\tilde{J}_{2}=3.4\cdot10^{-4}\;\mathrm{mol}/\mathrm{m}^2\mathrm{s}$. Similarly, for $E_m=0\;\mathrm{V}_{SHE}$ (pH$=3.7$, $\varphi=0.5\;\mathrm{mV}_{SHE}$) an approximation of the influx is given as $\tilde{J}_{2}=2\cdot10^{-5}\;\mathrm{mol}/\mathrm{m}^2\mathrm{s}$, and for $E_m=-0.5\;\mathrm{V}_{SHE}$ (pH=13, $\varphi=-0.03\;\mathrm{V}_{SHE}$) one reaches $\tilde{J}_{2}=1.1\cdot10^{-4}\;\mathrm{mol}/\mathrm{m}^2\mathrm{s}$. The global pH is the same used throughout the manuscript (pH=5). Relevant to the constant flux boundary condition, (ii), the values of hydrogen flux obtained using the electro-chemo-mechanical model after a time $t=10$ min. are $J_{10}=4.2\cdot10^{-5}\;\mathrm{mol}/(\mathrm{m}^2\mathrm{s})$, $J_{10}=1.5\cdot10^{-5}\;\mathrm{mol}/(\mathrm{m}^2\mathrm{s})$, and $J_{10}=1.5\cdot10^{-4}\;\mathrm{mol}/(\mathrm{m}^2\mathrm{s})$ for the $-0.5$, $0$, and $0.5\;\mathrm{V}_{SHE}$ metal potential cases, respectively.\\

The results obtained with each of the aforementioned modelling strategies are shown in Fig. \ref{fig:BC}. Two types of graphs are shown: the lattice hydrogen distribution ahead of the crack tip ($\theta=0^\circ$) after $t=10$ min. (left column) and the crack tip lattice hydrogen evolution as a function of time (right column). Consider first the results obtained for an applied potential $E_m=-0.5\;\mathrm{V}_{SHE}$, Figs. \ref{fig:BC}a and \ref{fig:BC}b. First, it can be readily observed that simply prescribing the initial pH of the electrolyte gives results that very significantly deviate from the reference case (the complete model). When the electrolyte is simulated, large changes in pH are observed near the metal surface and within the defect, with these changes limiting the reaction rates at the surface. However, by assuming a constant pH, this limiting effect is not present and as a result large amounts of $\mathrm{H}^+$ ions react at the surface, leading to an overprediction of the hydrogen uptake. A better approximation is obtained by considering the local pH, since this approach includes the aforementioned reaction rate limiting effects. However, this modelling strategy neglects changes in the electric potential of the electrolyte near the pit, leading to noticeable differences with the reference result. The accuracy compared to simply prescribing the global pH also improves when using the approximations $\tilde{J_1}$ (\ref{eq:mapping}) and $\tilde{J_2}$ (\ref{eq:env_simple}), as these take both the local pH and electrolyte potential into account. Within these, a better agreement is attained by making use of $\tilde{J_1}$, emphasising the role played by changes in surface coverage. A very good agreement can be obtained if the hydrogen influx is known due to a previous electro-chemo-mechanical simulation, as shown by the case $J_{10}$. However, one should note that this statement is only applicable for the case of $E_m=-0.5\;\mathrm{V}_{SHE}$ and the time scales considered. Specifically, one would expect predictions to worsen as the transient problem approaches the steady-state. In contrast, $\tilde{J}_2$ provides a worse approximation over short time scales but is likely to improve the steady-state prediction due to the inclusion of Tafel and Heyrovsky reactions. On the other side, prescribing a constant $C_L$, as commonly done in the literature, gives sensible results only if taking as input the outcome of a complete electro-chemo-mechanical analysis (such as the map provided in Fig. \ref{fig:hL_surfs}b) and only for the specific time instant considered (see Fig. \ref{fig:BC}b).\\
\FloatBarrier

For the $E_m=0\;\mathrm{V}_{SHE}$ and $E_m=0.5\;\mathrm{V}_{SHE}$ cases, prescribing a constant pH equal to the initial pH also produces poor results, as shown in Figs. \ref{fig:BC}c-f. In these cases, the actual $\mathrm{H}^+$ concentration is orders of magnitude higher than the initial one, such that the amount of absorbed hydrogen is significantly underestimated if the initial $\mathrm{H}^+$ concentration is considered. Prescribing the local pH provides better results for the neutral potential case but still results in negligible hydrogen being absorbed for the positive potential case. This is explained by the large differences in electrolyte potential observed in Fig. \ref{fig:Em_Phi_m05}, which accelerate the reaction rate of the hydrogen reactions but are not accounted for by solely prescribing a pH as boundary condition. As was also the case for the negative potential simulations, prescribing a constant lattice concentration results in a reasonable result at the time step this concentration is based on, but it does not capture the temporal behaviour correctly. Finally, the comparison of the results obtained using imposed hydrogen fluxes shows an offset between these and the reference results. This offset is caused by the inability of flux-based approaches to capture the impact that pH changes in time have. Specifically, the pH gradually decreases from the initial pH to the local environmental pH, causing the pH-dependent reactions to initially occur at a slow rate and only accelerate once the stable pH is reached, resulting in a lower amount of hydrogen initially entering the metal. However, after this initial period both the hydrogen flux based on results after 10 minutes ($J_{10}$) and the approximate hydrogen flux $\tilde{J}_1$ produce results that show similar gradients to the reference electro-chemo-mechanical model. The more simplified flux model $\tilde{J}_2$ also shows a similar trend but overpredicts the quantity of absorbed hydrogen as it neglects the hydrogen recombination reactions.

\section{Conclusions}
\label{sec:conclusion}

The amount of absorbed hydrogen in metals is a key input in hydrogen embrittlement predictions. However, its quantification remains a challenge. In this work, we have presented a generalised electro-chemo-mechanical model that enables quantifying hydrogen absorption for any choice of environment and sample/defect geometry. The model combines the simulation of ionic transport in an electrolyte with the diffusion of hydrogen within a deformable metal containing microstructural traps. At the interface between these two domains, electrochemical reactions are prescribed to relate the electrolyte pH and potential to the amount of hydrogen being absorbed into the metal. These elements are coupled, resulting in the first model that incorporates the physics governing electrolyte behaviour, hydrogen evolution and corrosion reactions, surface adsorption and stress-assisted hydrogen uptake and diffusion in a metal lattice. We numerically implement our theory and quantify hydrogen absorption as a function of the environment (bulk pH and applied potential), the fluid velocity and the crack and specimen dimensions. Furthermore, we postulate hypotheses and use them to present simplified versions of our model that enable quantifying the hydrogen influx from known local environmental conditions. Calculations are conducted to test these hypotheses and compare the predictions resulting from our simplified and generalised models to those obtained with the simplified boundary conditions commonly used in the literature, establishing regimes of validity. Our main findings are:
\begin{itemize}
    \item Hydrogen ingress shows significant sensitivity to changes in electrolyte potential and pH, despite these changes being neglected in existing models. Negative applied potentials reduce the $\mathrm{H}^+$ concentration and accelerate hydrogen reactions. The latter significantly enhances hydrogen uptake but the effect is limited due to the decrease in available $\mathrm{H}^+$ ions in the electrolyte. Positive potentials slow down hydrogen reaction kinetics but still lead to significant hydrogen uptake due to the associated reduction in pH. An intermediate regime exists where hydrogen ingress is minimised. 
    \item The fluid velocity has a minor influence on the hydrogen uptake ahead of cracks and pits. However, the bulk electrolyte potential and pH distributions are sensitive to the fluid velocity, and so is the hydrogen uptake at the exterior boundaries.
    \item Short and wide cracks/pits favour the diffusion of hydrogen ions into and out of the defect, with a stronger dependence on the defect length compared to its height. For sufficiently long cracks, this diffusion becomes severely limited, resulting in the pH becoming independent of the crack geometry. In contrast, the electrolyte potential exhibits a higher sensitivity to the defect geometry, even for long cracks.
    \item Neglecting electrolyte behaviour by defining the lattice hydrogen concentration at the surface based on the bulk pH introduces significant errors. Considering instead the local pH improves the accuracy of predictions but still shows deviations from the reference result, as changes in electrolyte potential are not accounted for. 
    \item Boundary conditions commonly used in hydrogen embrittlement models, such as prescribing a constant, pre-determined lattice concentration, result in significant deviations from the hydrogen absorption predicted by the complete electro-chemo-mechanical model.
    \item Environmental maps that relate the applied potential to the local (crack tip) pH and electrolyte potential, such as the one provided in Fig. \ref{fig:EnvironmentalMap}, can be used to determine the hydrogen influx in a relatively accurate manner, without the need to explicitly simulate electrolyte behaviour.
    \item While some simplifications provide relatively close estimates, there exists phenomena such as pH evolution that can only be captured with a complete electro-chemo-mechanical model. These phenomena lead to hydrogen uptake overpredictions when using simplified models that do not simulate the electrochemical behaviour of the electrolyte.
\end{itemize}

\noindent Moreover, maps are provided that enable readers to relate measurable environmental conditions (bulk pH, applied potential) to local environmental quantities (pH, electrolyte potential) and absorbed hydrogen. The model is also capable of predicting the influence of the surface condition but this is done through changes in the reaction rate constants and thus requires input from careful experimentation. Potential future extensions to the model include the use of kinetic trapping formulations \cite{McNabb1963,Turnbull2015}, the coupling with models that explicitly simulate the embrittlement process (e.g., through the use of phase field approaches \cite{TAFM2020c}) and incorporating the role of recombination poisons such as H2S. 

\section*{Acknowledgments}
\noindent Financial support through grant EP/V009680/1 (``NEXTGEM") from the Engineering and Physical Sciences Research Council (EPSRC) is gratefully acknowledged. Emilio Mart\'{\i}nez-Pa\~neda additionally acknowledges financial support from UKRI's Future Leaders Fellowship programme [grant MR/V024124/1].

\section*{Data availability}
\noindent The COMSOL physics builder model file incorporating the metal diffusion, interface reactions, and simplified boundary conditions is made freely available at \url{www.imperial.ac.uk/mechanics-materials/codes} and \url{www.empaneda.com}. Documentation is also provided, along with example files that enable to reproduce results shown in Section \ref{sec:bcs}.

\FloatBarrier
\appendix

\bibliography{references,library}

\end{document}